\documentclass[fleqn,usenatbib]{mnras}
\usepackage{newtxtext,newtxmath}
\usepackage[T1]{fontenc}
\usepackage{ae,aecompl}

\usepackage{graphicx}
\graphicspath{{Figures/}}
\usepackage{tabularx}
\newcolumntype{C}[1]{>{\centering\arraybackslash}p{#1}}
\usepackage{amsmath}	
\usepackage{amssymb}
\usepackage{wasysym}
\usepackage{times}

\usepackage[final]{changes}
\colorlet{Changes@Color}{red}

\title[SPMHD simulations of Structure Formation]{SPMHD simulations of Structure Formation}

\author[Barnes et al.]{David J. Barnes$^{1}$\thanks{Contact e-mail: \href{mailto:david.barnes@manchester.ac.uk}{david.barnes@manchester.ac.uk}},  Alvina Y. L. On$^2$, Kinwah Wu$^2$, Daisuke Kawata$^2$,
\\
$^{1}$Jodrell Bank Centre for Astrophysics, School of Physics and Astronomy, The University of Manchester, Manchester M13 9PL, UK\\
$^{2}$Mullard Space Science Laboratory, University College London, Holmbury St. Mary, Dorking, Surrey, RH5 6NT}

\date{Accepted XXX. Received YYY; in original form ZZZ}

\pubyear{2018}

\begin{document}
\label{firstpage}
\pagerange{\pageref{firstpage}--\pageref{lastpage}}
\maketitle

\begin{abstract}
The intracluster medium of galaxy clusters is permeated by $\mu\rm{G}$ magnetic fields. Observations with current and future facilities have the potential to illuminate the role of these magnetic fields play in the astrophysical processes of galaxy clusters. To obtain a greater understanding of how the initial seed fields evolve to the magnetic fields in the intracluster medium requires magnetohydrodynamic simulations. We critically assess the current Smoothed Particle Magneto-Hydrodynamics (SPMHD) schemes, especially highlighting the impact of a hyperbolic divergence cleaning scheme and artificial resistivity switch on the magnetic field evolution in cosmological simulations of the formation of a galaxy cluster using the N-body/SPMHD code \textsc{gcmhd+}+. The impact and performance of the cleaning scheme and two different schemes for the artificial resistivity switch is demonstrated via idealized test cases and cosmological simulations. We demonstrate that the hyperbolic divergence cleaning scheme is effective at suppressing the growth of the numerical divergence error of the magnetic field and should be applied to any SPMHD simulation. Although the artificial resistivity is important in the strong field regime, it can suppress the growth of the magnetic field in the weak field regime, such as galaxy clusters. With sufficient resolution, simulations with divergence cleaning can reproduce observed magnetic fields. We conclude that the cleaning scheme alone is sufficient for galaxy cluster simulations, but our results indicate that the SPMHD scheme must be carefully chosen depending on the regime of the magnetic field. 
\end{abstract}


\section{Introduction}
\label{sec:intro}
Observations of galaxy clusters show the presence of diffuse radio emission, unassociated with any obvious radio source, extending over $\text{Mpc}$ scales. The spectral index of this emission reveals the presence of relativistic electrons and large-scale magnetic fields pervading the intra cluster medium (ICM). Under the assumption of the minimum energy condition these magnetic fields are found to have a volume averaged strength of a few $\mu\text{G}$ \citep{GOV04}. Faraday rotation measurements of galaxies within a cluster infer a magnetic field that is not regularly ordered on large-scales, but that the cluster volume is a patch work of magnetic fields coherent on length scales of $5$ to $10\,\text{kpc}$ \citep{CAR02}. Observations indicate that the magnetic field strength radially decreases from a core at the centre of the galaxy cluster \citep{BON10}.

The origin of these large-scale magnetic fields is unclear, however, the proposed models fall in to two categories. These magnetic fields may be `primordial' in origin, initially generated in the very early Universe during inflation \citep{KAN11} or during matter phase transitions, such as the quark-hadron or electroweak transition \citep{KAH13}. These seed fields are then amplified via a turbulent dynamo mechanism driven by formation of large-scale structure. Primordial seed fields naturally have large correlation lengths and fill the volume of the Universe with magnetic fields, enabling them to account for the recent blazar observations that indicate cosmic voids contain $\text{kpc}$ scale coherent magnetic fields \citep{NER10,TAK13}. Current observations constrain Mpc coherent magnetic fields to have an amplitude less than $1.2\,\text{nG}$ \citep{PSH15}. However, many of these magnetogenesis mechanisms require physics beyond the standard model and it is unknown whether they are able to produce magnetic fields of sufficient amplitude to match the observations. Alternatively, large-scale magnetic fields may be seeded via plasma physics mechanisms during the formation of the first structures \citep{KUL97,ICH07,WID12}. These seed fields are then amplified by the collapse of structures and ejected by supernovae and active galactic nuclei (AGN) to `pollute' the large-scale structure of the Universe with magnetic fields. Seed magnetic fields generated by plasma processes are easily capable of producing the observed field amplitude, but struggle to account for the presence of strong magnetic fields in galaxies at high redshift and it is extremely difficult to reconcile them with the presence of coherent magnetic fields in cosmological voids.

Cosmological simulations \added{of structure formation that} follow\deleted{ing} the evolution of primordial seed fields have been performed using \added{a variety of techniques:} smoothed particle magnetohydrodynamics (SPMHD) \citep{Dolag1999,DOL09,BON11,STA13}, mesh codes \citep{BRU05,DUB08,MIN11,VAZ14} and adaptive mesh techniques \citep{MAR15,HOP15}. These simulations have shown that a seed field with an amplitude below the current observational upper limits of can be amplified to $\mu\text{G}$ amplitude by the formation of a galaxy cluster. The pollution of the ICM with magnetic energy via AGN injection has been studied by \citet{XU12} using an adaptive mesh refinement code. Assuming a model for the amount of magnetic energy injected by an AGN, they found that the ejected magnetic fields could reproduce the observed $\mu\text{G}$ amplitude magnetic field in a galaxy cluster. If the pollution occurs early enough during the formation of the cluster then the resulting magnetic field pervades the entire cluster volume. \added{In addition, simulations have examined the pollution of galaxy and cluster scale haloes with magnetic fields by supernovae seeding} \citep{DON09,Beck2013}. \added{They were able to reproduce the observed magnetic field within haloes, with galactic winds driving magnetic fields into the surrounding large-structure. However, they found that the filamentary structure of the universe is magnetized to a significantly lower level compared to simulations using a primordial seed magnetic field.}\deleted{performed SPMHD simulations of structure formation and coupled it with a semi-analytic prescription for magnetized galactic outflows, attributed to the starburst phase of galaxy evolution. The outflows pollute the ICM and are able to reproduce the observed magnetic field amplitude in galaxy clusters. However, they find that the filamentary structure of the universe is magnetized to a significantly lower level compared to simulations using a primordial seed magnetic field.}

Simulations have shown that the formation of structure erases any knowledge the magnetic field has about its initial seed \citep{Dolag2002,DOL09,BAR12,MAR15}. Therefore, to understand the origin of large-scale magnetic fields their co-evolution with structure formation must first be understood. A large dynamical range is essential in the numerical simulation in order to solve this problem. SPMHD simulations intrinsically provide this dynamic range, but they suffer from technical challenges, such as maintaining $\nabla\cdot\textbf{B}=0$, that can lead to the poor capture of dynamo processes and the incorrect evolution of a seed magnetic field. Recently, \citet{TRI12} have formulated a hyperbolic cleaning scheme to mitigate $\nabla\cdot\textbf{B}$ errors and \citet{TRI13} have suggested a new artificial resistivity switch. In this paper, we critically assess SPMHD formulations by implementing these new schemes in our N-body/SPMHD code, \textsc{gcmhd+} \citep{BAR12}. We critically assess different resistivity switches with and without hyperbolic divergence cleaning in cosmological simulations of the formation of a galaxy cluster. We highlight how the magnetic field evolves on cosmological timescales depending on the numerical scheme applied and demonstrate that the resultant magnetic field is very sensitive to the suppression of numerical divergence errors. We show that a primordial seed magnetic field is capable of reproducing observed magnetic field quantities in simulations with sufficient resolution.

The paper is organized as follows. In Section \ref{sec:method} we briefly lay out the additional implementations to our basic SPMHD implementation. In Section \ref{sec:itc} we show the impact of the additions to the SPMHD scheme have on idealized test cases. We then simulate the formation of a galaxy cluster in Section \ref{sec:MHDgcf} and look at how the chosen scheme affects the evolution of a primordial seed magnetic field during structure formation. Finally, in Section \ref{sec:concs} we give our conclusions.

\section{Numerical Method}
\label{sec:method}
We used the N-body/SPMHD code, \textsc{gcmhd+}, as the basis for this work. The hydrodynamic implementation and associated parameters are presented in \citet{KAW13} and the MHD implementation is presented in \citet{BAR12}. In brief, \added{we use a standard cubic spline kernel with $58$ neighbors. To resolve discontinuities in the fluid we include artificial viscosity, using the switch of} \citet{RosswogPrice2007} \added{, with $\alpha^{\mathrm{AV}}_{\mathrm{min}}=0.5$, to minimize viscosity where it is not required. We also include artificial conduction, including the switch of }\citet{Price2008}\added{, and employ the time-step limiter of }\citet{SaitohMakino2009}. The magnetic field is followed directly via the induction equation and it is allowed to act back on the fluid via the conservative magnetic stress tensor. The tensile instability, which results from the choice of the conservative stress tensor, is suppressed via the direct subtraction of any unphysical $\hat\beta\textbf{B}\left(\nabla\cdot\textbf{B}\right)/\rho$ force from the momentum equation \citep{BOR01}. We have changed from $\hat\beta=0.5$ in \citet{BAR12} to $\hat\beta=1$ to ensure that we avoid numerical artifacts \citep{TRI12}. An artificial resistivity scheme is used to resolve discontinuities in the magnetic field. The fastest magnetosonic wave speed is used to compute the time step and to control the artificial viscosity and resistivity schemes. However, there are still two outstanding issues with the SPMHD implementation. The first is the formulation of the switch used to control the application of resistivity at magnetic discontinuities. The second is maintaining the solenoidal condition of the magnetic field, $\nabla\cdot\textbf{B}=0$. Resolving these issues is critical to capturing the evolution of a seed magnetic field in a cosmological simulation accurately. To minimize the impact of these issues we have implemented additional algorithms in \textsc{gcmhd+}.

\subsection{Resistivity Switch Formulation}
\label{sec:rswitch}
\begin{figure*}
 \begin{centering}
 \includegraphics[width=\textwidth,keepaspectratio=true]{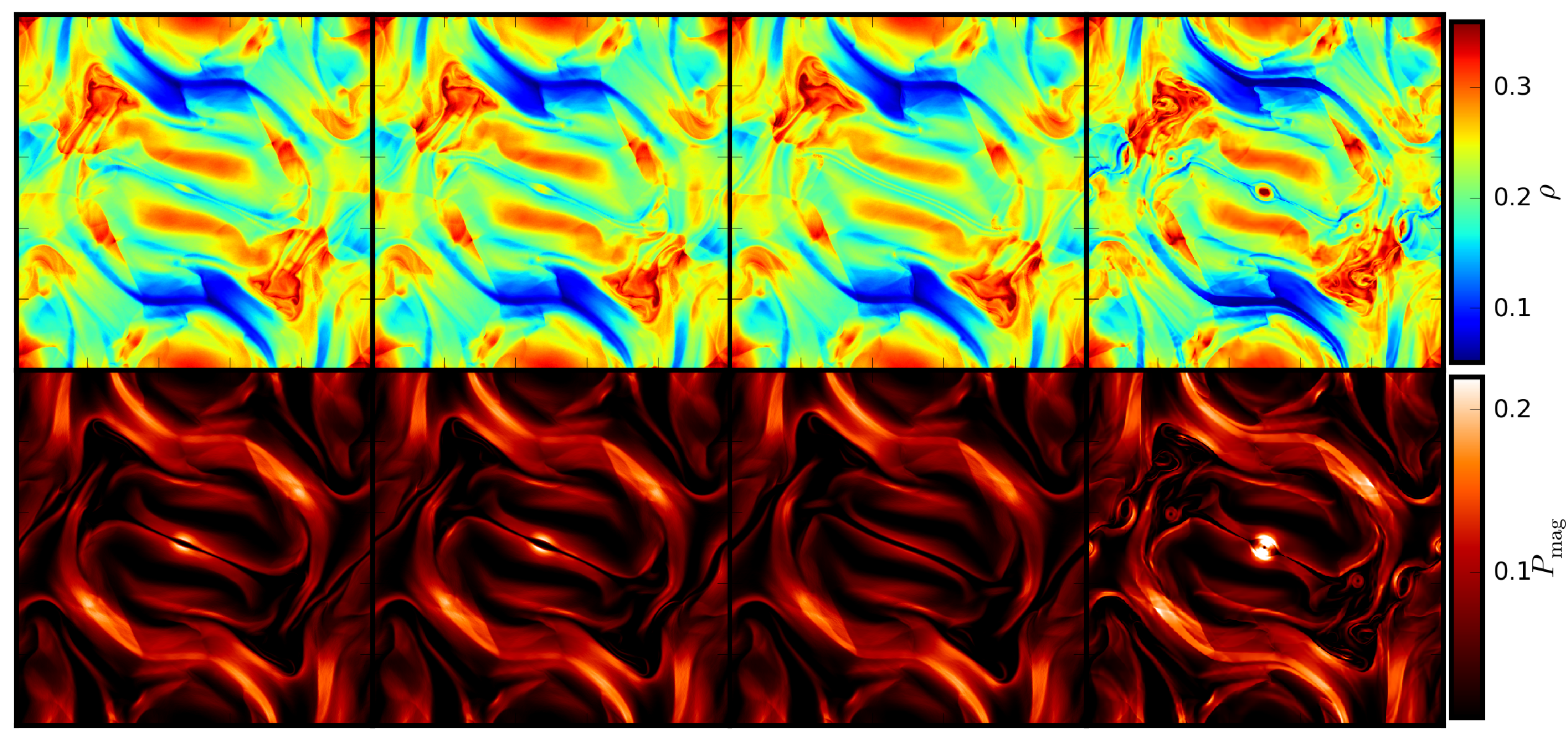}
 \caption{Density (top) and magnetic pressure (bottom) results of the Orszag-Tang vortex at $t=1.0$ for the standard (left), PM05 cleaned (centre left) and TP13 cleaned (centre right) \textsc{gcmhd+} implementations and the \textsc{athena} result (right). The large-scale features are similar, but all of the SPMHD simulations fail to fully capture the high density vortex that forms in the \textsc{athena} result.}
 \label{fig:OTden_pmag}
 \end{centering}
\end{figure*}

Artificial resistivity is a dissipative term included in SPMHD simulations to ensure that discontinuities in the magnetic field are correctly captured. The resistivity dissipates the magnetic field and smooths the discontinuity so that it is resolved. This allows the correct pre- and post-shock values of the magnetic field to be obtained. In many astrophysical systems, such as in galaxies and galaxy clusters, the magnetic Reynold's number is very high and, therefore, it is critical to minimize the resistive dissipation where it is not required. In \citet{BAR12}, \textsc{gcmhd+} used the switch proposed by \citet{PRI05} (henceforth referred to as the PM05 switch) with $\alpha^{B}_{\text{min}}= 0.0$ to reduce dissipation away from discontinuities. However, \citet{PRI12} showed that even with this switch there was still significant unwanted dissipation in their simulations. \citet{TRI13} showed that this switch fails to capture discontinuities in the weak field regime due to $\alpha^B$, and hence the applied resistivity, remaining low in the presence of large discontinuities due to its dependence on the magnitude of the magnetic field. This is problematic for cosmological MHD simulations as structure formation generates many shock and the magnetic field is always in the weak field regime.

\citet{TRI13} proposed an alternate resistivity switch (henceforth referred to as the TP13 switch) where $\alpha^B$ is independently set for each particle directly to the dimensionless quantity
\begin{equation}
 \alpha^B_i=\frac{h_i|\nabla\textbf{B}_i|}{|\textbf{B}_i|}\:,
\end{equation}
where $h_i$ is the smoothing length of particle $i$, $\nabla\textbf{B}$ is the $3\times3$ gradient matrix of the magnetic field $\textbf{B}$ and $\alpha^B_i$ is limited such that $\alpha^B_i\in\left[0,1\right]$. The individual components of the matrix are calculated via the standard SPH operator
\begin{equation}
 \nabla\textbf{B}_i\equiv\frac{\partial\textbf{B}^k_i}{\partial x^l_i}\approx-\frac{1}{\Omega_i\rho_i}\sum_jm_j\left(\textbf{B}^k_i-\textbf{B}^k_j\right)\nabla^l_iW_{ij}(h_i)\:,
\end{equation}
where $\rho_i$ is the density at particle $i$, $m_j$ is the mass of the particle $j$, $\Omega_i$ accounts for the variable smoothing length terms and $W$ is the SPH smoothing kernel, which is the cubic spline kernel in this implementation. The 2-norm is used to calculate the norm of $\nabla\textbf{B}$ 
\begin{equation}
 |\nabla\textbf{B}|\equiv\sqrt{\sum_k\sum_l\left|\frac{\partial\textbf{B}^k_i}{\partial x^l_i}\right|^2}\:.
\end{equation}
We have implemented this proposed resistivity switch in \textsc{gcmhd+}. In section \ref{sec:itc} we examine how the choice of resistivity switch effects the result of idealized test cases and in section \ref{sec:MHDgcf} we examine the impact of both switches in an MHD cosmological simulation.

\begin{figure*}
 \begin{centering}
 \includegraphics[width=\textwidth,keepaspectratio=true]{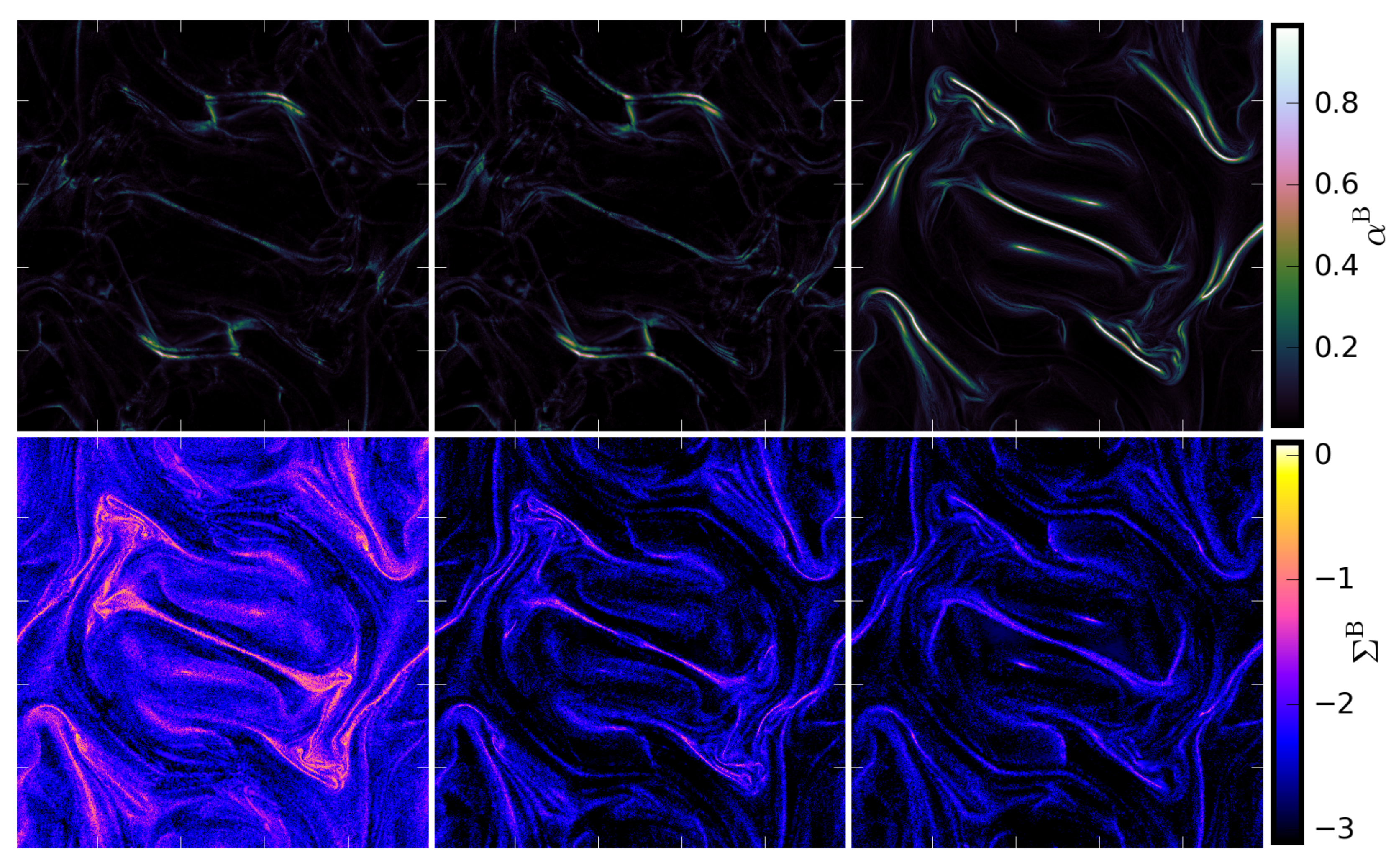}
 \caption{Resistivity coefficient $\alpha^{B}$ (top) and divergence error $\Sigma^{B}$ (bottom) results of the Orszag-Tang vortex at $t=1.0$ for the standard (left), PM05 cleaned (centre) and TP13 cleaned (right) implementations. The TP13 switch demonstrates better shock tracking and the hyperbolic cleaning scheme produces an order of magnitude reduction in the divergence error.}
 \label{fig:OTres+divb}
 \end{centering}
\end{figure*}

\subsection{Hyperbolic Cleaning Scheme}
\label{sec:hcs}
In MHD simulations using an SPH based code the numerical integration of the magnetic field leads to non-zero field divergence and the violation of the solenoid condition. The violation of this condition can lead to spurious forces parallel to the magnetic field, potentially resulting in numerical instabilities, and to the poor capture of dynamo processes, so it is critical to minimize any violation. We implement a hyperbolic cleaning scheme, originally proposed by \citet{DED02}, in \textsc{gcmhd+} to minimize the build of numerical divergence of the magnetic field. This scheme has been widely used in Eulerian codes \citep{AND06,MIG10} and has been adapted for SPMHD simulations \citep{TRI12,STA13} and moving mesh simulations \citep{PAK11}. The hyperbolic cleaning scheme evolves an additional scalar field $\psi$, which is coupled to the magnetic field and leads to an additional term in the induction equation of the form
\begin{equation}\label{eq:scalarcoup}
 \left(\frac{d\textbf{B}}{dt}\right)^{\text{clean}}=-\nabla\psi\:.
\end{equation}
The scalar field evolves according to
\begin{equation}\label{eq:psievo}
 \frac{d\psi_i}{dt}=-v^2_{M,i}\nabla\cdot\textbf{B}_i-\frac{\psi_i}{\tau}\:,
\end{equation}
where $\psi_i$ is the value of the scalar field at particle $i$, $v_{M,i}$ is the fast magnetosonic wave speed at $i$ and $\tau$ is the damping timescale, which is equal to
\begin{equation}
 \frac{1}{\tau}=\frac{\sigma v_{M,i}}{h_i}\:,
\end{equation}
where $\sigma$ is a free parameter that determines the strength of the damping. By combining equations (\ref{eq:scalarcoup}) and (\ref{eq:psievo}), in the co-moving frame of the wave, a damped wave equation is obtained that demonstrates how the scheme works. The divergence of the magnetic field is propagated away from a source like a wave and the wave is damped by the diffusion term. The strength of the damping applied to the divergence wave is determined by the choice of $\sigma$. It has been suggested by \citet{MIG10} that $\sigma\in[0,1]$ and \citet{TRI12} suggest that $0.2<\sigma<0.3$ for {$2$D} tests and $0.8<\sigma<1.2$ for {$3$D} tests, while \citet{STA13} find a value of $\sigma=4.0$ produced optimal results. After testing $\sigma$ values in the range $\in[0.2,5.0]$ with a suite of standard MHD tests and in cosmological MHD simulations and we found a value of $\sigma=1.0$ produces the best compromised results for all simulations. All of the simulations presented in this paper use $\sigma=1.0$.

When hyperbolic cleaning is included as part of the MHD equations, \citet{TRI12} argue that an additional term in equation (\ref{eq:psievo}) is required to ensure total energy conservation. If this term is included then equation (\ref{eq:psievo}) becomes
\begin{equation}
 \frac{d\psi_i}{dt}=-v^2_{M,i}\nabla\cdot\textbf{B}_i-\frac{\psi_i}{\tau}-\frac{\psi_i\nabla\cdot\textbf{v}_i}{2}\:,
\end{equation}
where $\textbf{v}_i$ is the velocity of particle $i$. We ran all of our test and cosmological simulations with and without this additional  energy conservation term. Although we found that the inclusion of the term had negligible effect on the results of the simulations, it is included in all the results presented in this paper. We calculate divergence of the magnetic field using the SPH `difference' operator \citep{PRI12} and we measure the `divergence error' via 
\begin{equation}
 \Sigma^{\rm{B}}=\log_{10}\left(\frac{h_i\nabla\cdot\textbf{B}_i}{|\textbf{B}_i|}\right)\:.
 \label{eq:diverr}
\end{equation}

\section{Idealised Test Cases}
\label{sec:itc}
\begin{figure*}
 \begin{centering}
 \includegraphics[width=\textwidth,keepaspectratio=true]{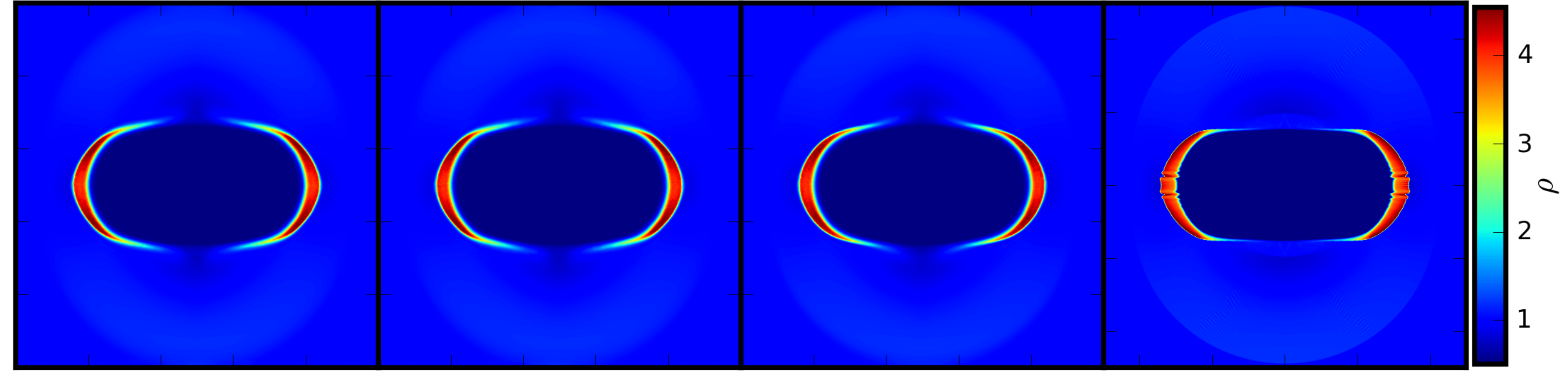}
 \caption{Density result for the magnetized blast wave test at $t=0.03$ standard (left), PM05 cleaned (centre left) and TP13 cleaned (centre right) implementations of \textsc{gcmhd+} and the \textsc{athena} result (right). Neglecting the numerical artifact in the shock front of the mesh code, the results are in good agreement.}
 \label{fig:MBWden}
 \end{centering}
\end{figure*}

To ensure that the code, with the additional resistivity switch and hyperbolic cleaning scheme, was producing reliable results we thoroughly tested it with a range of standard idealized MHD test cases. For many of these tests no analytic solution exists and in these cases we compared the result produced by \textsc{gcmhd+} against the result produced by the \textsc{athena} grid code \citep{STO08}. Although we have examined the performance of the code for a full range of MHD test cases \added{(see Appendix \ref{sec:AppST})}, the idealized test cases presented in this section demonstrate the impact of the additional schemes presented in the previous section, i.e. changing the resistivity switch or including the hyperbolic cleaning scheme.

\subsection{Orszag-Tang Vortex}
\label{sec:OTv}
The Orszag-Tang compressible vortex \citep{ORZ79} test produces several interacting shock fronts that then transitions to turbulence and is a common test of many MHD implementations. We set up an ideal gas, with $\gamma=5/3$, within a {$2$D} periodic square box of length unity. The gas is given an initial velocity defined by $v_x=-\sin\left(2\pi y\right)$, $v_y=\sin\left(2\pi x\right)$ and $v_z=0$. The gas is embedded with an initial magnetic field of $B_x=-B_0\sin\left(2\pi y\right)$, $B_y=B_0\sin\left(4\pi x\right)$ and $B_z=0$, where $B_0=1/\sqrt{4\pi}$. The initial pressure is set to $P=\gamma B_0^2$ and the initial density of the gas is set to $\rho=\gamma P$. All of the results presented in this section use $1024\times1180$ particles arranged initially in a hexagonal lattice configuration. The complex dynamics that result from this set up provide an excellent test of the chosen resistivity switch and the hyperbolic cleaning scheme.

\begin{center}
 \begin{table}
  \begin{tabularx}{\columnwidth}{l|C{1.4cm}|C{0.9cm}|C{3.1cm}}
  \textsc{gcmhd+} & Artificial & Switch & Hyperbolic cleaning \\
  scheme & Resistivity & & scheme \\ 
  \hline
  Standard & Varying & PM05 & $-$ \\
  Control & None & $-$ & $-$ \\
  Constant & Constant & $-$ & $-$ \\
  Cleaned & None & $-$ & $\surd$ \\
  PM05 cleaned & Varying & PM05 & $\surd$ \\
  TP13 cleaned & Varying & TP13 & $\surd$ \\
  \hline
  \end{tabularx}
  \caption{Table summarizing the different configurations of \textsc{gcmhd+} used in the simulations presented in this work.}
  \label{tab:scheme}
 \end{table}
\end{center}

To clearly demonstrate the effect of adding the cleaning scheme and changing the formulation of the resistivity switch we ran this test with three different configurations of the code. The `standard' implementation uses the PM05 resistivity switch and does not use the cleaning scheme (the code as presented in \citet{BAR12}), the `PM05 cleaned' implementation uses the PM05 resistivity switch and includes the hyperbolic cleaning scheme and the `TP13 cleaned' implementation uses the TP13 resistivity switch and includes the cleaning scheme. A summary of all implementations used throughout this work is given in Table \ref{tab:scheme}. We evolved the initial set up using these implementations and the \textsc{athena} code to $t=1.0$ and the density and magnetic pressure results are shown in Fig. \ref{fig:OTden_pmag}. 

The density results are qualitatively comparable to the \textsc{athena} result, with all of the SPMHD implementations showing some slight intrinsic smoothing of features. The \textsc{athena} result shows the formation of a very high density vortex in the centre of the test. This feature is reproduced to some extent by the implementations using the PM05 switch, but this feature is absent from the result produced by the TP13 cleaned implementation. The magnetic pressure results produced by the implementations show the same global features as the \textsc{athena} result. The failure to accurately capture the very dense central vortex feature produces a significant reduction in the central magnetic pressure of the results produced by \textsc{gcmhd+}. The use of the TP13 switch further suppresses the central magnetic pressure.

The Orszag-Tang vortex's increasingly complex shock fronts provide a good test of a resistivity switch's ability to track the shock fronts and the cleaning scheme's ability to maintain the solenoidal condition of the magnetic field. Fig. \ref{fig:OTres+divb} shows the applied resistivity parameter and divergence error, as defined by equation (\ref{eq:diverr}), for the three implementations at $t=1.0$. The TP13 switch accurately tracks the shock fronts as they become more complex, applying significantly more resistivity at shock fronts and less away from them. The PM05 resistivity switch is not as good at tracking the complex shock fronts and applies resistivity more broadly across the simulation area. Comparing the cleaned implementations to the standard implementation shows that the inclusion of the hyperbolic cleaning scheme produces an order of magnitude reduction in the divergence error of the magnetic field. Inclusion of the cleaning scheme keeps the divergence of the magnetic field at a few percent of the total magnetic field amplitude throughout the simulation. The choice of resistivity switch has a minimal impact of the divergence error of the magnetic field, but it is slightly lower on average for the TP13 resistivity switch.

\begin{figure*}
 \begin{centering}
 \includegraphics[width=\textwidth,keepaspectratio=true]{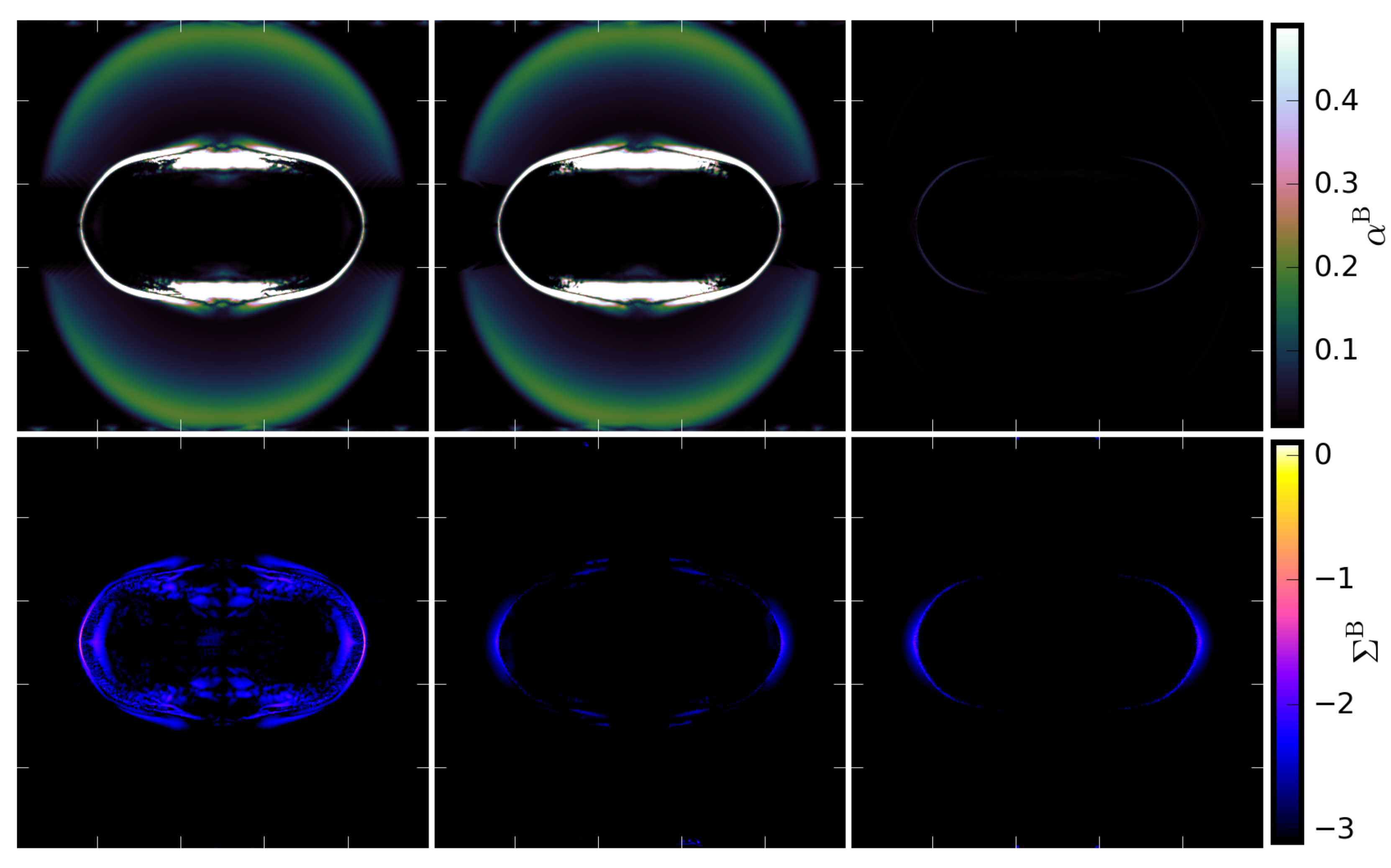}
 \caption{Resistivity coefficient $\alpha^B$ (top) and divergence error $\Sigma^B$ (bottom) results for the magnetized blast wave test at $t=0.03$ for the standard (left), PM05 cleaned (centre) and TP13 cleaned (right) implementations. The TP13 switch only applies resistivity to the leading edge of the shock fronts. The hyperbolic cleaning scheme produces a significant reduction in divergence error throughout the simulation.}
 \label{fig:MBWres_divb}
 \end{centering}
\end{figure*}

\subsection{2D MHD Blast Wave}
\label{sec:MHDblast}
The blast wave test is the explosion of over-pressured gas particles in a static, magnetized background medium and is standard test of MHD codes. We set up $512\times590$ particles on a hexagonal lattice in a {$2$D} periodic square box of length unity. The gas has a density $\rho=1$, a pressure $P_0=1$, an adiabatic index $\gamma=1.4$, and is at rest. A hot disc of radius $r_0=0.125$, centred on the middle of the domain, is set up such that particles that fall inside the radius of the disc have their pressure increased by a factor of 100, so that $P_{disc}=100$. All gas particles are embedded with an initially homogeneous magnetic field that has an amplitude of $B_x=10$, $B_y=0$ and $B_z=0$ and is orientated in the {$x$-direction} only. As the initial set up is evolved a strong outward traveling shock wave develops, but it is constrained perpendicular to the magnetic field. This test provides an excellent test of the effectiveness of the different resistivity switches and the hyperbolic cleaning scheme in the strong shock and low beta regimes. We evolved the initial set up using the three SPMHD implementations and compare to the result produced by \textsc{athena}. Fig. \ref{fig:MBWden} shows the density result at $t=0.03$ and the results produced by the SPMHD configurations are qualitatively identical to the reference result produced by \textsc{athena}, neglecting the numerical artifact present in the mesh result.

To demonstrate the impact of the resistivity switch and the cleaning scheme, Fig. \ref{fig:MBWres_divb} shows the resistivity parameter, $\alpha^B$, and the divergence error, $\Sigma^B$, at $t=0.03$ \added{for the standard, PM05 cleaned and TP13 cleaned implementations}. The TP13 resistivity switch results in significantly lower values of $\alpha^B$ throughout the simulation, with only non-negligible values occurring at the leading edge of the shock fronts. In contrast the PM05 resistivity switch results in values of $\alpha^B$ approaching unity at the shocks and where the shocks are suppressed perpendicular to the magnetic field. In addition, use of the PM05 switch leads to $\alpha^B$ increasing behind the fast magnetoacoustic wave as it propagates outwards. The lower values of $\alpha^B$ produced by the TP13 switch and resulting smaller smoothing of the magnetic field mean that this test is effectively run at higher resolution compared to using the PM05 switch. Independent of the chosen resistivity switch, the inclusion of the hyperbolic cleaning scheme produces an order of magnitude reduction in the divergence error throughout the simulation. The divergence of the magnetic field is maintained to at most a few percent of the total amplitude of the magnetic field.

\begin{figure*}
 \begin{centering}
 \includegraphics[width=\textwidth,keepaspectratio=true]{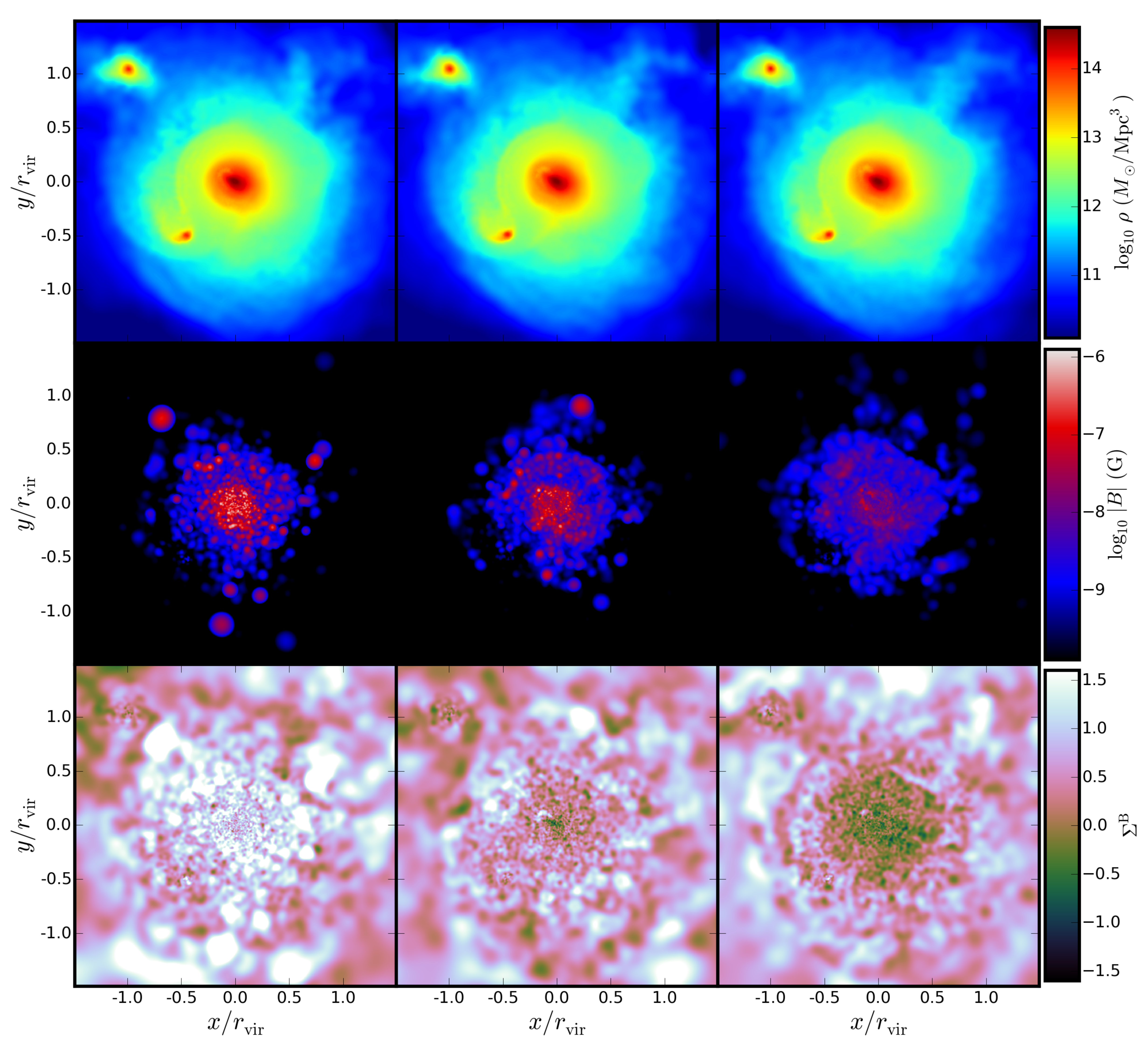}
 \caption{Density (top row), magnetic field amplitude (centre row) and divergence error (bottom row) $xy$ slices at $z=0$ for the galaxy cluster simulated using the control (left), constant (centre) and standard (right) implementations. Strong magnetic field in the control and constant simulations is associated with large divergence errors.}
 \label{fig:GCmap1}
 \end{centering}
\end{figure*}

Including the hyperbolic cleaning scheme significantly surpasses the growth of the divergence of the magnetic field and has negligible impact on the result of the test simulations. The TP13 resistivity switch produces a significant improvement in the artificial resistivity scheme's ability to track the development of complex shock fronts, but it applies more resistivity at shock fronts leading to some features being smoothed out. We now investigate the impact of different numerical schemes, including hyperbolic cleaning and resistivity switch formulation, on magnetic fields in cosmological simulations.

\section{Cosmological MHD Simulations of Galaxy Cluster Formation}
\label{sec:MHDgcf}
In the paradigm of hierarchical structure formation, galaxy clusters form from the collapse and coalescence of many smaller objects to become the most massive gravitationally bound objects in the Universe. The merging of smaller objects and the accretion of material produces shocks and induces turbulence that has been observed in the ICM \citep{SCH04}. Shocks and turbulence in the ICM will amplify an initial seed magnetic field and merging will generate bulk motions that re-distributes the magnetic field throughout the cluster volume. To use observations of large-scale magnetic fields to infer their origin or their impact on the astrophysical processes within the cluster, it is crucial to model the formation and evolution of the galaxy cluster over cosmic time using numerical MHD simulations. We simulate the formation of a galaxy cluster with a primordial seed magnetic field embedded within the gas. We examine the evolution of the seed field due to the formation of the cluster and demonstrate the sensitivity of the magnetic field to the choice of numerical scheme, i.e. chosen resistivity scheme and suppressing the growth of the divergence of the magnetic field.

We assumed a flat $\Lambda\text{CDM}$ cosmological model with cosmological parameters taken from the WMAP 5 year data release, such that $h=0.72$, $\Omega_m=0.26$, $\Omega_b=0.044$, $\sigma_8=0.8$ and $n_s=0.96$ \citep{KOM09}. At $z=60$, a low resolution $200\,\text{Mpc}$ dark matter only spherical volume was seeded with density perturbations using the {\tt GRAFIC2} code \citep{BER01}. These initial conditions were then evolved to $z=0$ and a central halo with a mass of $M_{200}=1.5\times10^{14}\,\mathrm{M}_{\astrosun}$ was selected, where $M_{200}$ corresponds to the mass inside the radius at which the average density is equal to two hundred times the critical density of the Universe. This halo was then refined twice, where particles that fell within eight and four times \deleted{the} $r_{200}$ were replaced with $8\times$ and $64\times$, respectively, lower mass particles to produce a ``zoomed'' simulation of the formation of the halo. The refinement radii were set so that resulting zoomed halo is free from contaminating higher mass particles out to approximately $3\times r_{200}$. The highest refined particles each had a mass of $m_{DM}=6.20\times10^{8}\,\mathrm{M}_{\astrosun}$, with a corresponding softening length $\epsilon=6.2\,\text{kpc}$.

\begin{figure*}
 \begin{centering}
 \includegraphics[width=\textwidth,keepaspectratio=true]{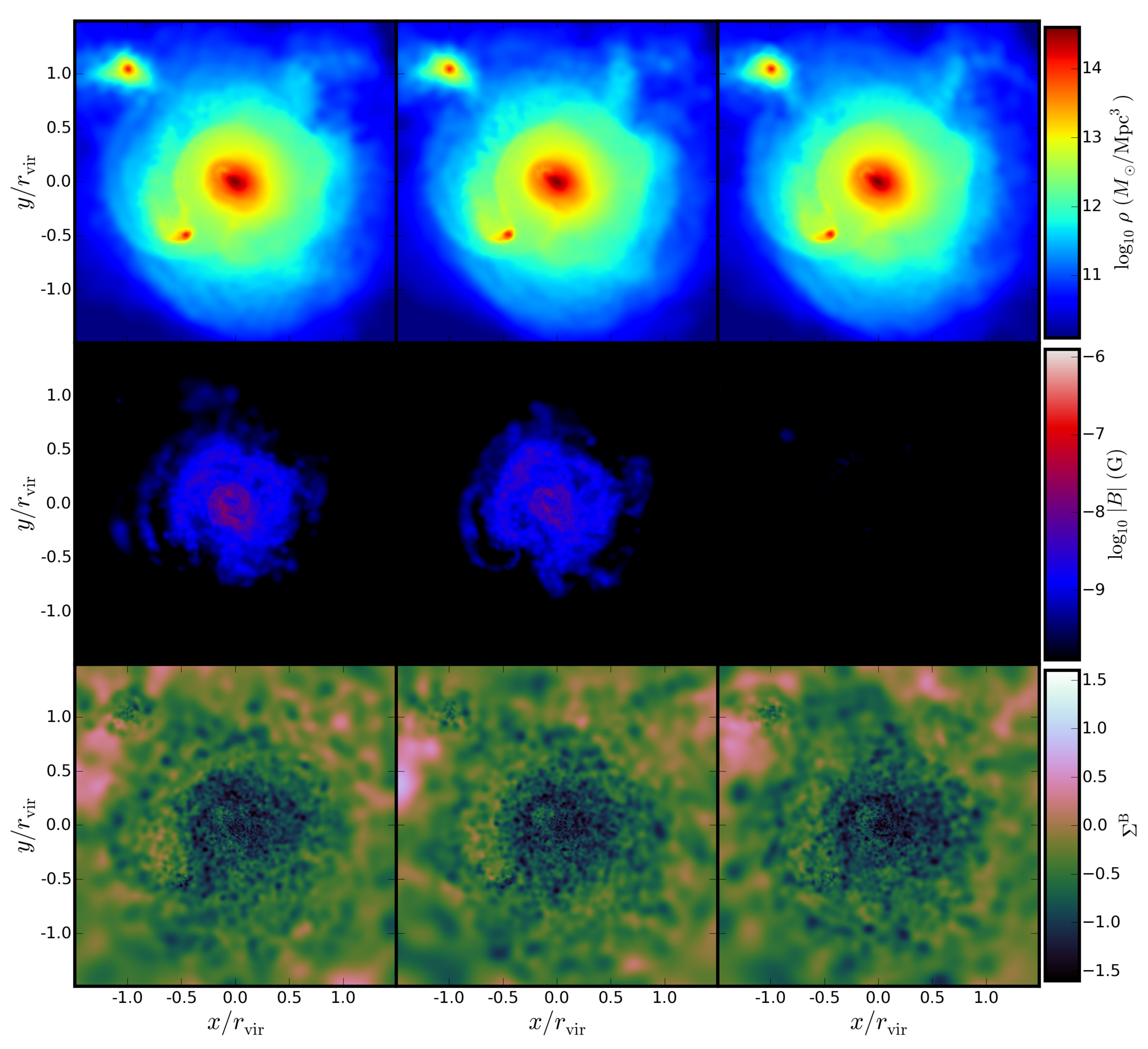}
 \caption{Same as Fig. \ref{fig:GCmap1} except that the galaxy cluster has been simulated using the cleaned (left), PM05 cleaned (centre) and TP13 cleaned (right) implementations. The hyperbolic cleaning scheme significantly reduces the divergence error within the cluster volume.}
 \label{fig:GCmap2}
 \end{centering}
\end{figure*}

To these initial conditions gas particles, each with a mass of $m_g=1.26\times10^8\,\mathrm{M}_{\astrosun}$, were added to the highest resolution region. The gas particles were embedded with an initially homogeneous seed magnetic field that had an amplitude of $10^{-11}\,\text{G}$. As the final magnetic field has been shown to be independent of the spectral properties of the seed field \citep{DOL09,BAR12,MAR15}, the initial seed field was orientated in the {$x$-direction} only, to ensure that it was divergence free. To focus on the how the seed field evolves during the formation of the cluster, the simulations were purely adiabatic in nature and neglected additional physical processes such as radiative cooling, star formation and feedback from stars and AGN.

To examine the impact of chosen numerical scheme on the evolution of the seed magnetic field, the initial conditions were run using six configurations of \textsc{gcmhd+}. In addition to the standard, PM05 cleaned and TP13 cleaned implementations present in Section \ref{sec:OTv}, the simulation was run with the `control', `constant' and `cleaned' implementations, see Table \ref{tab:scheme}. The control implementation simply evolved the induction equation and did not use an artificial resistivity scheme or the hyperbolic cleaning scheme. The constant implementation applied a constant level of resistivity equal to $\eta=6.4\times10^{27}\,\text{cm}^2\text{s}^{-1}$ to each gas particle at every time step, similar to \citet{BON11,BEC13}, and did not use the hyperbolic cleaning scheme. The cleaned implementation used the hyperbolic cleaning scheme, but did not use an artificial resistivity scheme. These were chosen so that impact of different schemes could be assessed independently of each other. All six simulations used the same initial conditions and were evolved to $z=0$.

Figs. \ref{fig:GCmap1} and \ref{fig:GCmap2} show density, magnetic field amplitude and divergence error slices through the $xy$ plane of the final cluster for the six simulations. The choice of numerical MHD scheme has negligible impact on the formation of the cluster and all six implementations produce a cluster with a mass of $M_{200}=1.51\times10^{14}\,\mathrm{M}_{\astrosun}$ and a similar morphology. This is expected as the magnetic energy density is orders of magnitude smaller than kinetic energy of the forming cluster and the magnetic field is effectively passively frozen into the gas. However, the chosen scheme has a significant impact on the topology and amplitude of the magnetic field. The amplitude of the magnetic field varies from $10^{-6}\,\text{G}$ in the control and constant implementations to $10^{-10}\,\text{G}$ in the TP13 cleaned implementation. Those simulations that reach $\mu\text{G}$ amplitude have a magnetic field that is a patchwork of strong magnetic field in a background magnetic field that is significantly weaker. These simulations have no mechanism for \replaced{minimising the violation of $\nabla\cdot\textbf{B}=0$}{enforcing the solenoidal condition of the magnetic field} and the stronger magnetic field amplitude is associated with large divergence errors. Those simulations that include the hyperbolic cleaning scheme have significantly lower divergence errors, but the magnetic field amplitude reaches a maximum of $\sim 10^8\,\text{G}$. Therefore, the amplitude of the magnetic field at $z=0$ is dependent on the amplitude of the numerical \added{and unphysical} divergence of the magnetic field.

\begin{figure}
 \begin{centering}
 \includegraphics[width=\columnwidth,keepaspectratio=true]{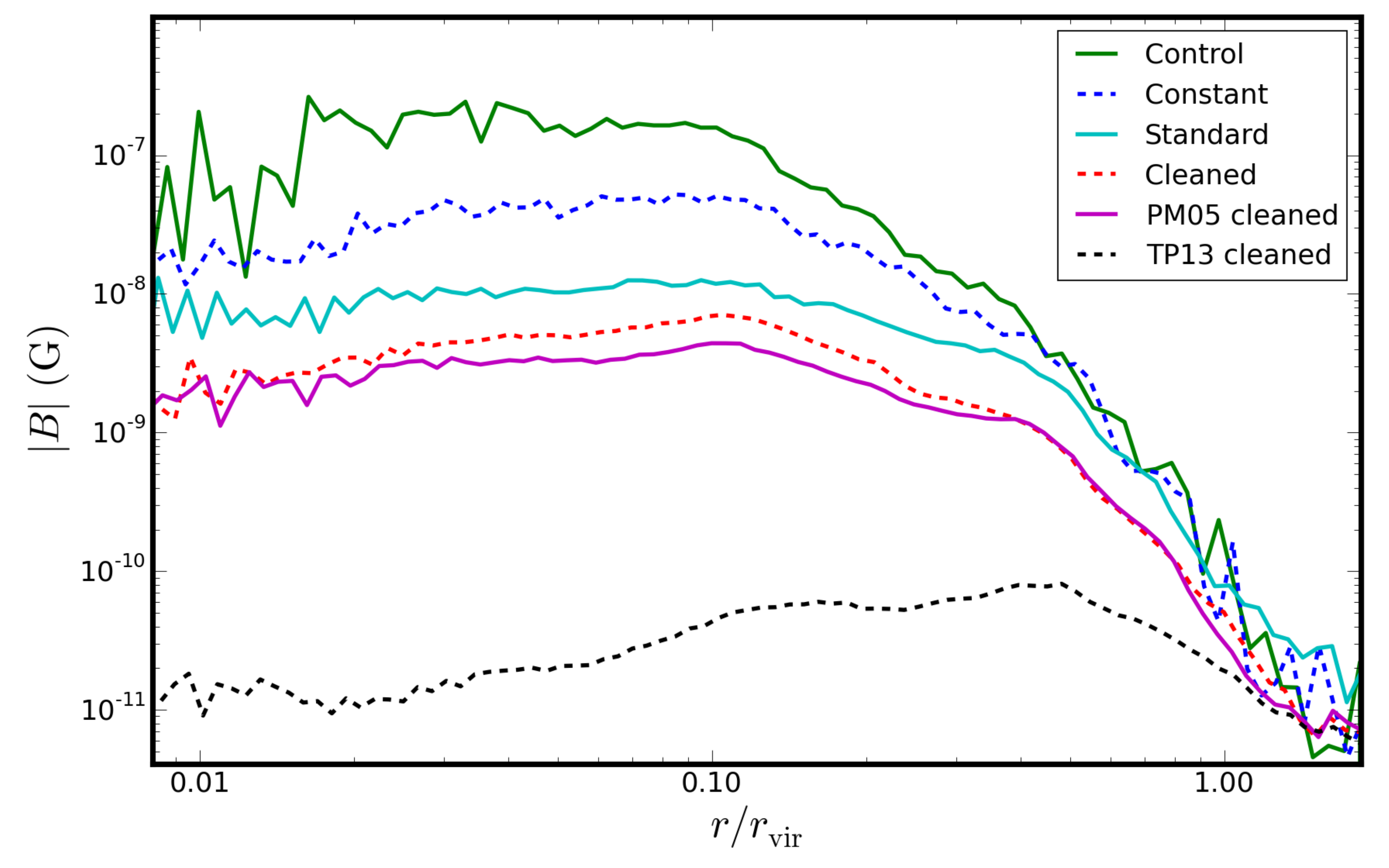}
 \includegraphics[width=\columnwidth,keepaspectratio=true]{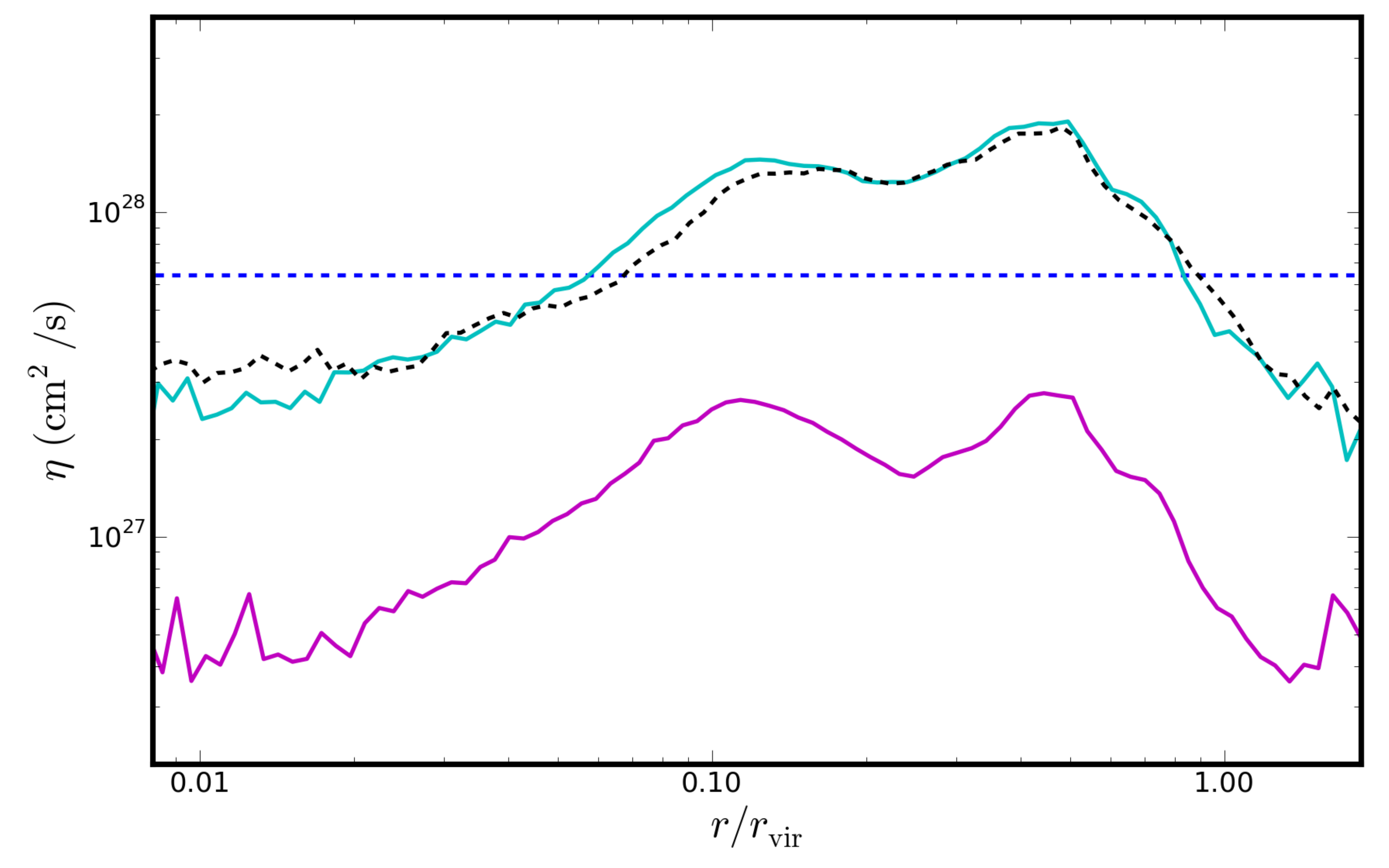}
 \includegraphics[width=\columnwidth,keepaspectratio=true]{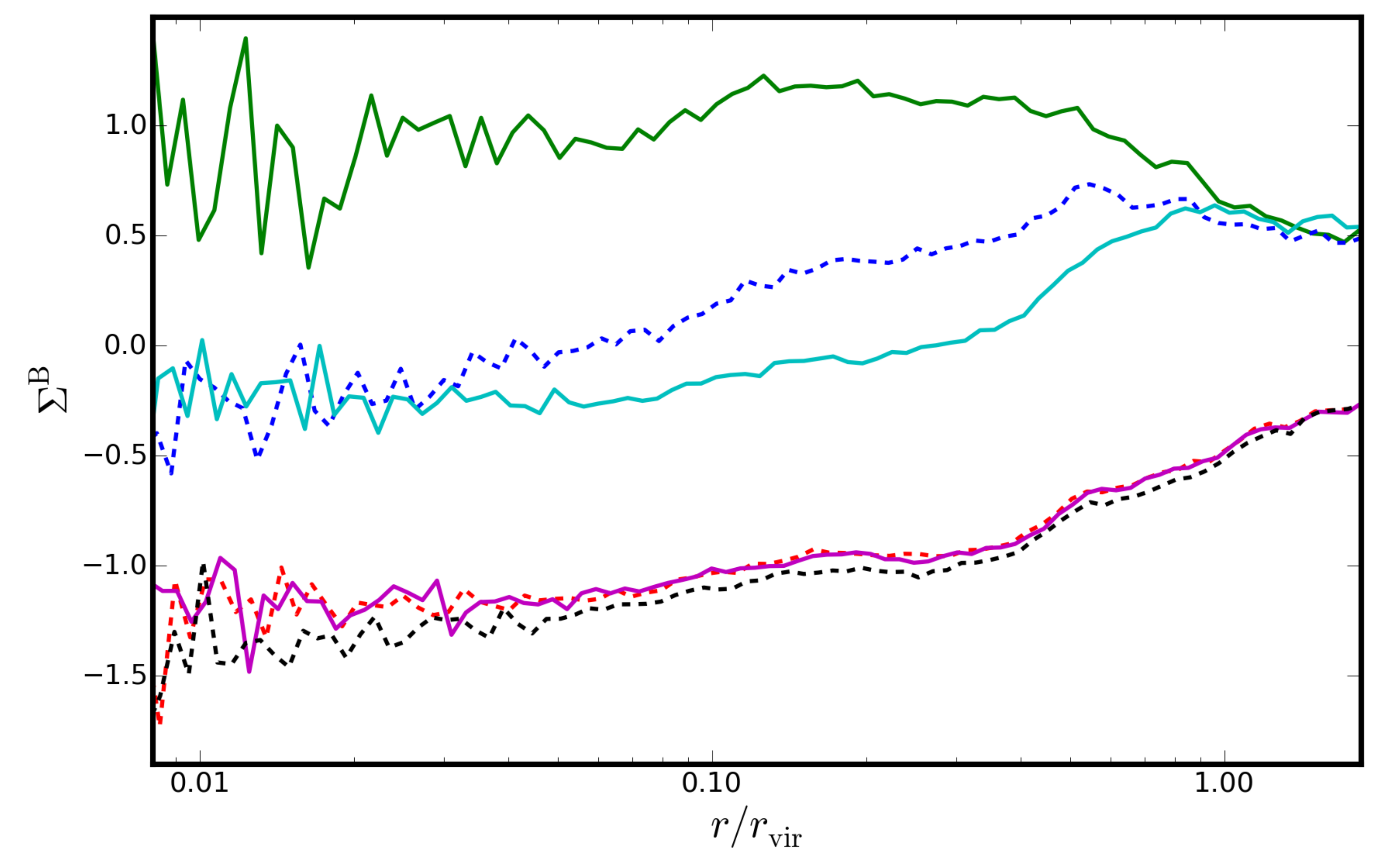}
 \caption{Magnetic field amplitude (top), resistivity (middle) and divergence error (bottom) radial profiles of the galaxy cluster at $z=0$ for the six different implementations. The TP13 resistivity switch is too strong for cosmological simulations.}
 \label{fig:GCprofiles}
 \end{centering}
\end{figure}

Fig. \ref{fig:GCprofiles} shows the magnetic field amplitude, average applied resistivity and divergence error radial profiles of the six simulations. With the exception of the TP13 cleaned \replaced{simulation}{cluster} all of the clusters reproduce the observed magnetic field profile with a central core that declines steeply with increasing radius \citep{BON10}. The resistivity profiles demonstrate that the TP13 resistivity switch applies an order of magnitude more resistivity to the particles than the PM05 switch, which applies resistivity of a similar level to the constant implementation. This remains the case when $\alpha^{B}$ is limited to $[0,0.1]$, as suggested by the authors for more realistic simulations. The resulting cluster magnetic field has been shown in previous numerical work to be very sensitive to the level of applied resistivity \citep{DOL09,BAR12}. Therefore, the TP13 switch is too strong for cosmological simulations and suppresses the amplification of the magnetic field. A comparison of constant and standard runs demonstrates a switch to control the application of resistivity is required because the resistivity can vary by an order of magnitude throughout the cluster volume. The importance of minimising the growth of the divergence of the magnetic field is shown by comparing the standard and PM05 cleaned simulations. Suppressing its growth produces a factor of $\sim5$ reduction in the level of resistivity applied. Therefore, the unphysical numerical divergence of the magnetic field is controlling the application of the resistivity and in turn is controlling the evolution of the seed magnetic field.

The divergence error profile of the control simulation shows that failing to include an artificial resistivity scheme or the hyperbolic cleaning scheme leads a situation where the amplitude of the divergence of the magnetic field is \replaced{a factor $\sim10$}{ten times} larger than the amplitude of the magnetic field at $z=0.0$. The implementation of an artificial resistivity scheme lead to \added{a} reduction in the divergence of the magnetic field. This is because the resistivity damps the true magnetic field and the unphysical divergence of the magnetic field. However, simply increasing the resistivity to damp the divergence will also damp the true magnetic field. The inclusion of the hyperbolic cleaning leads to a significant decrease in the divergence error, with the chosen resistivity scheme now producing a negligible impact on the divergence error of the PM05 cleaned and TP13 cleaned simulations. The inclusion of the hyperbolic cleaning scheme suppresses the unphysical divergence of the magnetic field to a few percent of the magnetic field amplitude throughout the cluster volume.

\subsection{Seed Field Evolution}
\begin{figure}
 \begin{centering}
 \includegraphics[width=\columnwidth,keepaspectratio=true]{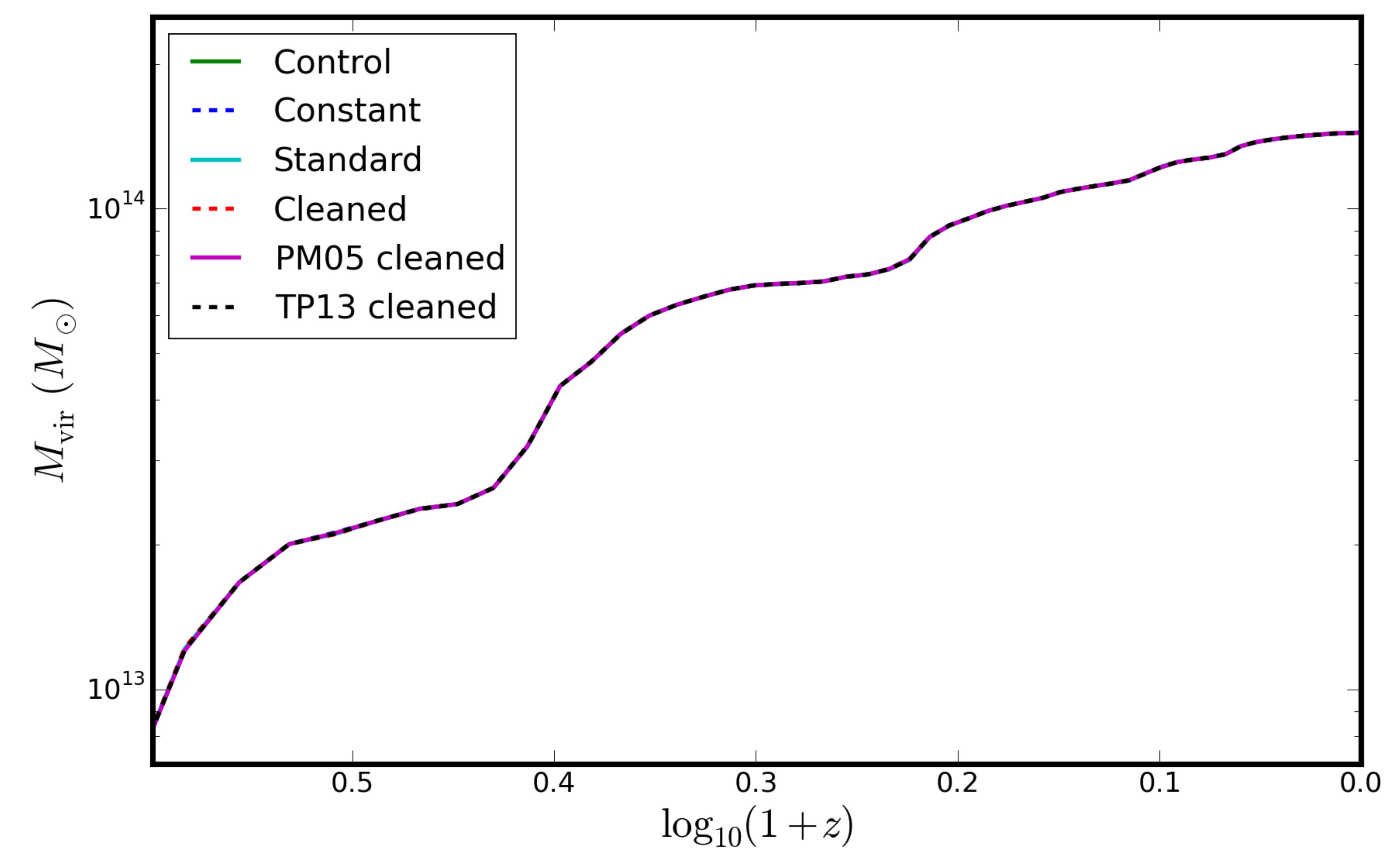}
 \includegraphics[width=\columnwidth,keepaspectratio=true]{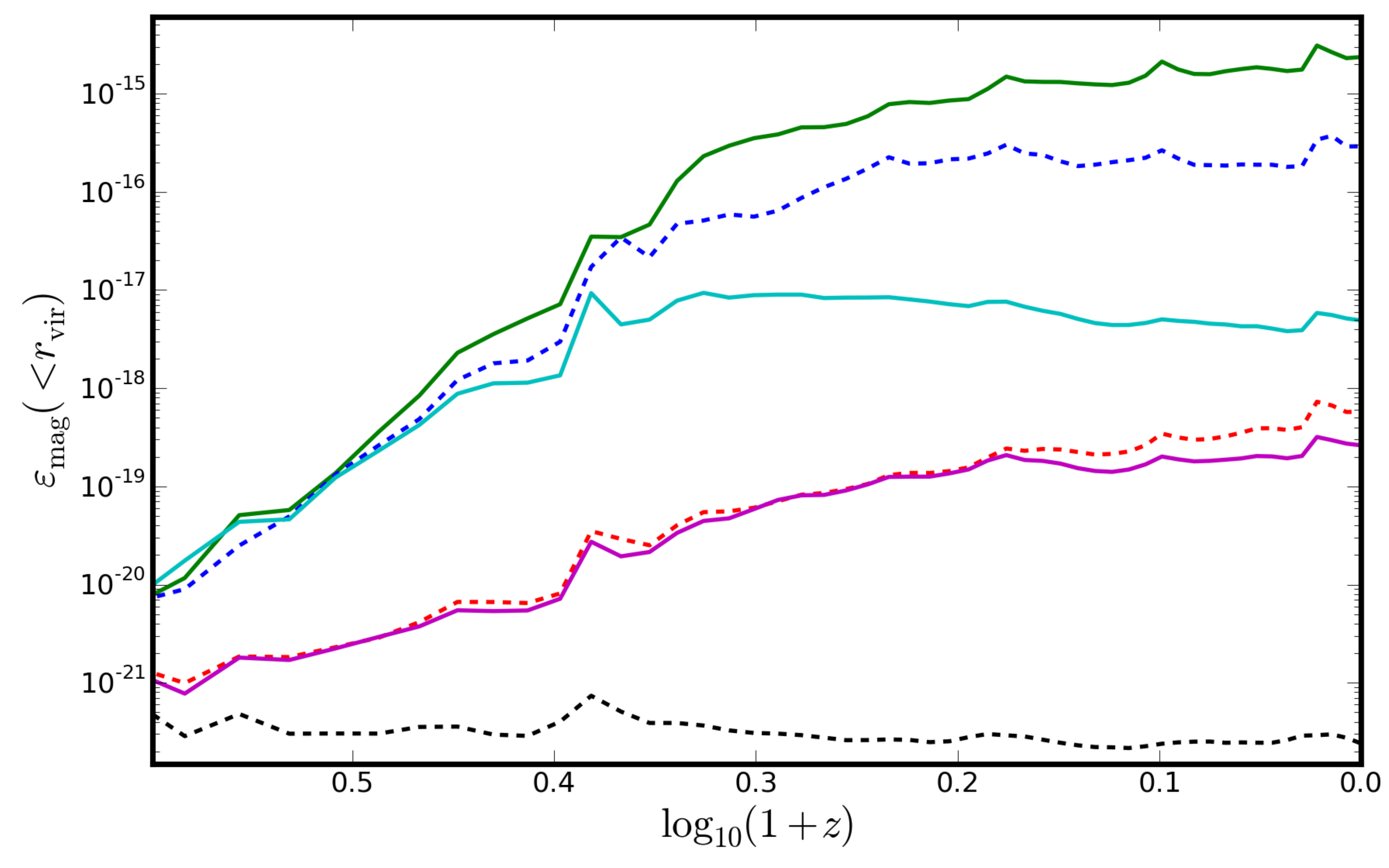}
 \includegraphics[width=\columnwidth,keepaspectratio=true]{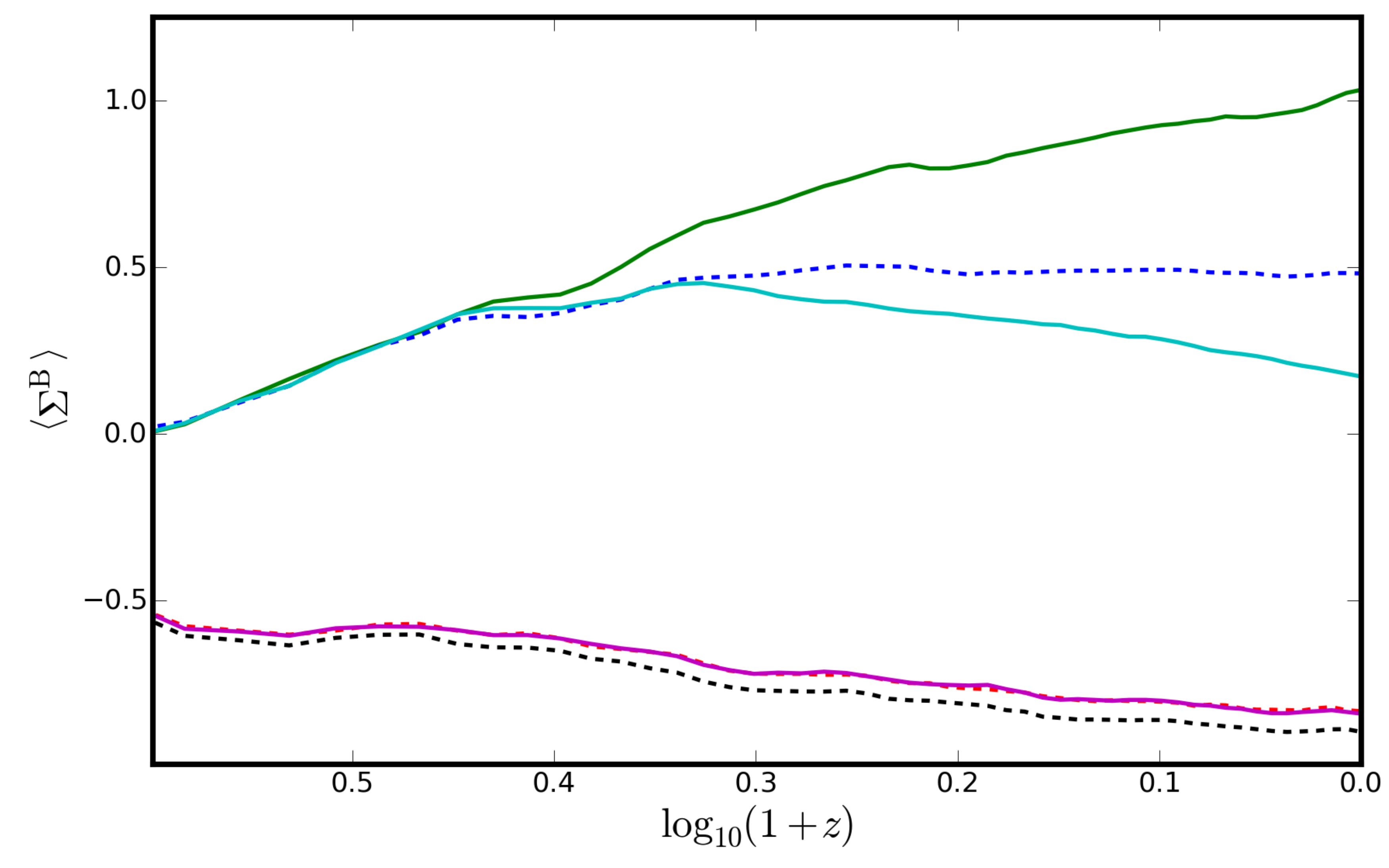}
 \caption{Plots showing the evolution of $M_{200}$, the average magnetic energy density and the average divergence error of the galaxy cluster between $z=3$ and $z=0$ for the six different implementations.}
 \label{fig:EVOprofiles}
 \end{centering}
\end{figure}

To further understand the impact of the chosen numerical scheme on the evolution of the seed magnetic field during structure formation we follow the most massive progenitor of the cluster from $z=3.0$ to $z=0.0$. Fig. \ref{fig:EVOprofiles} shows the $M_{200}$ of the progenitor, its magnetic magnetic energy density inside $r_{200}$, $\varepsilon_{\text{mag}}$, and its average divergence error inside $r_{200}$ as a function of $\log_{10}(1+z)$. All of the simulations show the same growth of $M_{\mathrm{vir}}$ and the chosen MHD scheme has negligible effect on the growth of the cluster. \added{As seen in previous work, for example }\citet{Roettiger1999} \replaced{ , the}{The} magnetic energy density of the cluster increases during major merging events, such as $x=0.41\:(z=1.57)$ and $x=0.21\:(z=0.62)$. The collapse of material provides the energy to drive the dynamo mechanism that amplifies the magnetic field, while the merging events are features on trend of increasing magnetic energy with decreasing redshift. The turbulent motions induced by mergers and accretion of material are the dominant mechanism for amplifying the seed magnetic field.

There is a correlation between the average divergence error of the cluster and its rate of increase of the magnetic energy density. The small-scale turbulent folding and shearing motions induced by the collapse of structure amplify the true magnetic field. However, they amplify the unphysical divergent magnetic field as well. Hence, the amplification of the seed field is then dependent on both the true and unphysical magnetic field. Therefore, it is critical to suppress the \added{growth of the} divergence error in the simulations to avoid spurious amplification of the seed magnetic field due to the presence of the unphysical divergence of the magnetic field. This is demonstrated in the standard simulation, once the resistivity becomes strong enough to reduce the average divergence error of the cluster the rate of increase of the magnetic energy flattens. The hyperbolic cleaning scheme produces the most significant reduction in the divergence error and so including it in the numerical scheme is critical to capturing the true evolution of the seed magnetic field.

However, those simulation\added{s} that include the hyperbolic cleaning scheme fail to produce cluster magnetic fields with the observed $\mu\text{G}$ amplitude. To fully capture the amplification of the magnetic field the simulations must be able to resolve the small-scale velocity field. We examined the effect of resolution on the amplification of the seed magnetic field by running the simulations with increased resolution.

\subsection{Higher Resolution Simulations}
To produce a higher resolution simulation of the formation of the galaxy cluster the refinement of the highest resolution level was increased. Instead of replacing each particle that fell within $4\times r_{200}$ with $8$ times lower mass particles they were replaced with $64$ times lower mass particles. This produced initial conditions with increased resolution, such that the highest resolution dark matter particles each had a mass of $m_{DM}=7.8\times10^7\,\mathrm{M}_{\astrosun}$, with a corresponding softening of $\epsilon=2.6\,\text{kpc}$, and each gas particle had a mass of $m_g=1.6\times10^7\,\mathrm{M}_{\astrosun}$. The gas particles were embedded with an initially homogeneous magnetic field that had an amplitude of $10^{-11}\,\text{G}$. The field was orientated entirely in the {$x$-direction} to ensure that is was initially divergence free.

\begin{figure*}
 \begin{centering}
 \includegraphics[width=\textwidth,keepaspectratio=true]{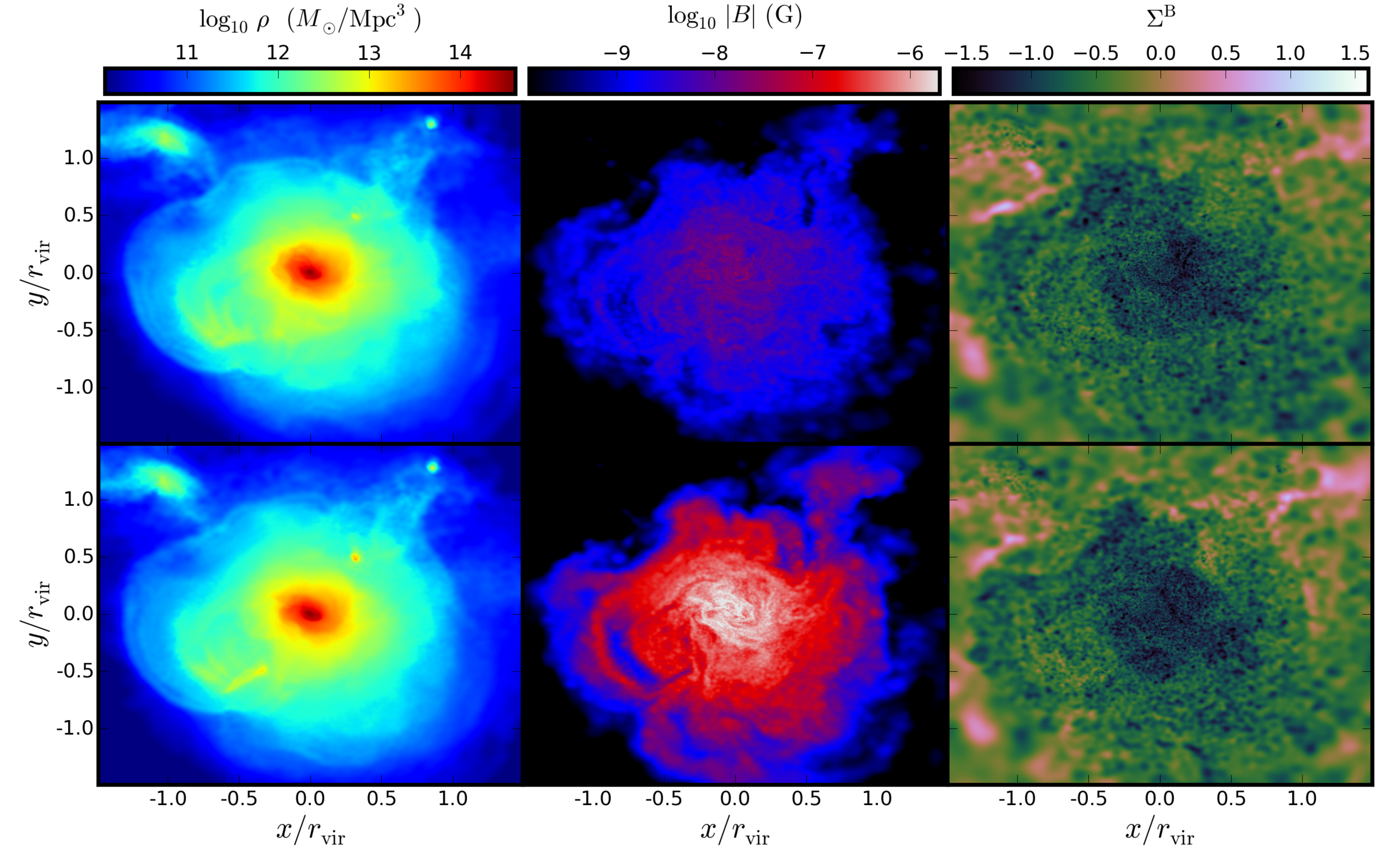}
 \caption{Density (left), magnetic field amplitude (centre) and divergence error(right) $xy$ slices of the higher resolution galaxy cluster at $z=0$ for the PM05 cleaned (top) and cleaned (bottom) implementations.}
 \label{fig:HRmap}
 \end{centering}
\end{figure*}

The higher resolution initial conditions were evolved using the cleaned and PM05 cleaned implementations of \textsc{gcmhd+}, because the lower resolution simulations in the previous section showed that these two schemes were a compromise between suppressing the growth of the numerical divergence of the magnetic field, but not suppressing the amplification of the magnetic field. Slices through the $xy$ plane of the density, magnetic field amplitude and divergence error quantities at $z=0.0$ for the two clusters are shown in Fig. \ref{fig:HRmap}. The density slices show that the two different schemes have negligible impact on the formation and global hydrodynamic properties of the cluster. Both clusters include the hyperbolic cleaning scheme and the divergence of the magnetic field is maintained to a few percent of the amplitude of the magnetic field amplitude throughout the cluster. However, the amplitude of the magnetic field is significantly different, with the cleaned implementation producing a cluster with a central magnetic field amplitude of $1.5\,\mu\text{G}$, in agreement with the observations, while the PM05 cleaned cluster has a central amplitude of $0.016\,\mu\text{G}$. 

\begin{figure}
 \begin{centering}
 \includegraphics[width=\columnwidth,keepaspectratio=true]{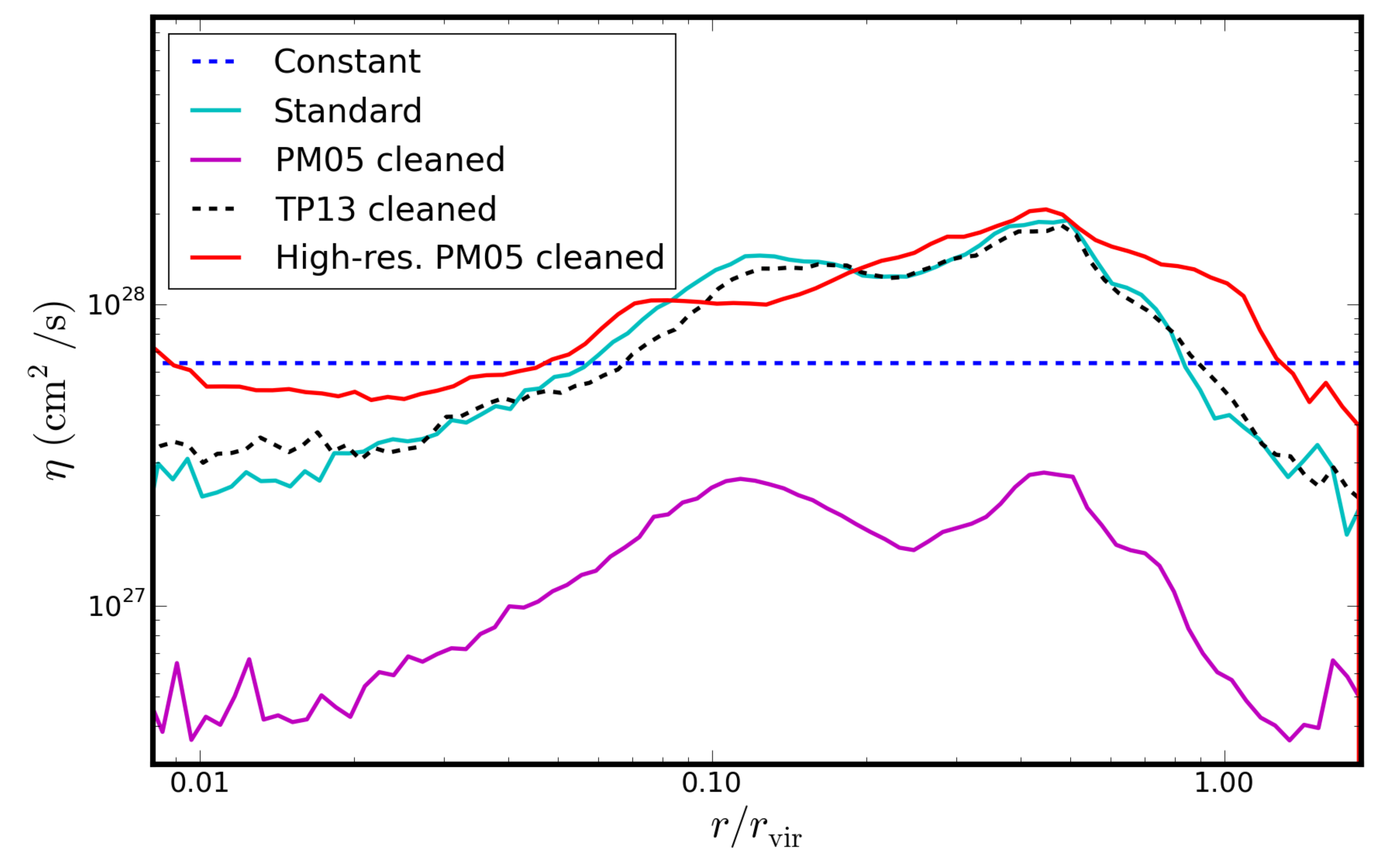}
 \caption{Resistivity radial profile for the standard resolution simulations and the higher resolution PM05 cleaned simulation. The increase in resolution produces a significant increase in resistivity throughout the cluster volume.}
 \label{fig:HRresistivity}
 \end{centering}
\end{figure}

The difference between the implementations is the use of the artificial resistivity scheme. Fig. \ref{fig:HRresistivity} shows the resistivity applied by the different implementations in the standard and higher resolution simulations. A higher resolution simulation resolves smaller scale structures and fluctuations, which results in greater amplification of the seed magnetic field. However, this increase in small-scale structure causes the PM05 resistivity scheme to apply a factor \replaced{$\sim10$}{of ten} greater resistivity throughout the volume of the cluster. This increase in the strength of the resistivity applied suppresses the amplification of the magnetic field. Therefore, the cluster simulated using the cleaned implementation produces \added{a} cored magnetic field with $\mu\text{G}$ amplitude, in agreement with the observations, while the cluster simulated using the PM05 cleaned implementation produces a cored magnetic field with $100\times$ smaller amplitude. Artificial resistivity is included in these simulations to ensure that magnetic discontinuities are numerically resolved, otherwise resistivity would not be required as the resistive timescale of a cluster is significantly greater than the age of the Universe. However, these simulations have an initially continuous magnetic field \replaced{with}{, and} no sources or sinks of magnetism and do not resolve the turbulence scale of the cluster. Therefore, any discontinuity in the magnetic field will be associated with discontinuities in other hydrodynamic properties. These will be smoothed out and resolved by other schemes, such as artificial viscosity, and so an artificial resistivity scheme is not required at the resolution of the simulations presented in this paper.

\subsection{X-ray and Radio emission properties}
\deleted{The higher resolution cleaned simulation produces a cluster with a total mass of about $10^{14}\,\mathrm{M}_{\astrosun}$ and a central peak magnetic field strength of $\approx\,2.5\;\mu\text{G}$, which are consistent with values derived from observations and previous numerical simulations, see e.g. Thierbach et al. (2003); Bonafede et al. (2010), and also reviews by Govoni \& Feretti (2004); Brunetti \& Jones (2014). }Observationally, \deleted{only} $20 - 30\%$ of massive, X-ray luminous clusters are found to host diffuse, Mpc-scale radio haloes of low surface brightness, typically of several $\mu\text{Jy}\,\text{arcsec}^{-2}$ at $1.4\,\text{GHz}$ \citep[e.g.][]{FER12}. \added{These clusters are normally associated with signs of recent merger activity \citep{Cassano2010}}. Those clusters studied in detail, e.g.\, Coma, A2255, A2744 \& A2163, show similar morphological features in their radio and X-ray images \citep[e.g.][]{LIA00, FER01, GOV01A, GOV01B, BAC03, GOV04}. Upon a local point-by-point comparison between the radio surface brightness and X-ray photon count rate, the correlation is found to be adequately described by a power law \citep{FER01, GOV01A}.

\added{The high-resolution cleaned simulation produces a cluster with a central peak magnetic field of $\approx\,2.5\;\mu\text{G}$, which is in reasonable agreement with values derived by previous numerical work} \citep[e.g.][]{BON11} \added{and those found observationally} \citep{GOV04,BON10,BRU14}. \replaced{Although the cluster is an order of magnitude smaller in mass than clusters with observed radio haloes, we now simulate its X-ray and radio morphology and examine if a power-law naturally arises.}{Here we examine whether the power law relation arises naturally from MHD simulations with certain sensible assumptions.} We use the cleaned implementation of \textsc{gcmhd+} and consider \replaced{the}{a} higher resolution simulation \deleted{to generate a magnetised cluster} at redshift $z = 0$. We then calculate the corresponding X-ray and radio emissions from the cluster. Each synthetic map/image of the cluster spans roughly $2.9\,\mathrm{Mpc}\,\times\,2.9\,\mathrm{Mpc}$ with a depth of $2.9\,\mathrm{Mpc}$ and is divided into $256 \times 256$ square pixels.  

\added{Clusters are X-ray luminous primarily due to thermal free-free radiation emitted from hot ($10^{5.2}\,\text{K} \leq T \lesssim 10^{7}\,\text{K}$) and diffuse ($10^{-6}\,\text{cm}^{-3} \lesssim n_e \lesssim 10^{-2}\,\text{cm}^{-3}$) intra-cluster gas}. The continuum X-rays are thermal bremssthrahlung radiation, with an emissivity given by 
\begin{equation}
\mathtt{j}_{\text X}=\frac{{\rm{d}}W}{{\rm{d}}t ~{\rm{d}}V} =\left(\frac{2 \pi k_{\text B} T_{\mathrm{e}}}{3 m_{\mathrm{e}}}\right)^{1/2}\frac{2^{5} \pi e^{6}}{3 h m_{\mathrm{e}} c^{3}}Z^{2}n_{\mathrm{e}}n_{\mathrm{i}}\bar{g}_{\text B}\:,
\label{eq:emisX}
\end{equation}
in erg\,s$^{-1}$\,cm$^{-3}$ \citep{RYB79}, where $m_{\mathrm{e}}$ is the electron mass, $e$ is the electron charge, $c$ is the speed of light, $k_{\text B}$ is the Boltzmann constant and $h$ is the Planck constant.
$n_{\mathrm{e}}$ and $T_{\mathrm{e}}$ are the number density and temperature of the thermal electrons, respectively. We set the velocity-averaged Gaunt factor $\bar{g}_{\text B}$ to $1$ without losing generality. 
Assuming a fully ionised hydrogen plasma, we set $Z = 1$ and $n_e = n_i$. Since the plasma is fully ionised at $T = 10^{5.2}$\,K, we only take into account of emission from particles with temperatures higher than such. 
In addition to that, we also only consider emission from regions below the critical density to form stars, i.e. $\rho \le 0.01$\,M$_{\astrosun}$\,pc$^{-3}$ \citep{COX06, GEN12}. We calculate the bolometric X-ray intensity by integrating the emissivity over the interval d$s$ along the line-of-sight
\begin{equation}
I_{\text X}\propto\int {\rm{d}}s ~n_{\mathrm{e}}^2 ~T_{\mathrm{e}}^{1/2}\:, 
\label{eq:IX}
\end{equation}
in erg\,s$^{-1}$\,cm$^{-2}$.

The radio emission is synchrotron radiation from energetic electrons gyrating around the cluster magnetic fields. We assume relativistic electrons with a power-law energy spectrum
\begin{equation}
n_\text{re}(\gamma) ~d\gamma = \mathcal{C} \gamma^{-p} ~d\gamma \: ,
\end{equation}
where $p$ is the electron energy spectral index and $n_\text{re}(\gamma) ~d\gamma$ is the number density of  relativistic electrons in the energy range of $\gamma$ and $\gamma+d\gamma$. Thus, the energy density of the relativistic electrons is 
\begin{equation}
\mathcal{E}_\text{re} = \mathcal{C} (m_{\mathrm{e}}c^{2})\int_{1}^{\infty}d\gamma \  \gamma^{1-p}  \: .
\end{equation} 
Assuming that $\mathcal{E}_\text{re}$ is $1 \%$ of the thermal energy density  \citep{KOT11,GEN12}, the normalisation constant is 
\begin{equation}
\mathcal{C} = 3 \times 10^{-3} \left(\frac{n_{\mathrm{e}} k_{\text B} T_{\mathrm{e}}}{m_{\mathrm{e}} c^{2}}\right) \:.
\end{equation}
As such, the specific radio emissivity is 
\begin{multline}
j_{\text R} = \frac{{\rm{d}}W}{{\rm{d}}t ~{\rm{d}}V ~{\rm{d}}\nu} = \frac{\sqrt{3} e^{3} \mathcal{C} B_{\perp}}{m_{\mathrm{e}} c^{2} (p+1)}\left(\frac{m_{\mathrm{e}} c 2 \pi \nu}{3 e B_{\perp}} \right)^{-(p-1)/2} \\
\times\Gamma \left(\frac{p}{4} + \frac{19}{12} \right)\Gamma \left(\frac{p}{4} - \frac{1}{12} \right)\,,
\end{multline}
in erg\,s$^{-1}$\,cm$^{-3}$\,Hz$^{-1}$ \citep{RYB79} at frequency $\nu$, where $B_{\perp}$ is the magnetic field component perpendicular to the line-of-sight and $\Gamma(...)$ is the Gamma function. We \replaced{assume}{consider} $\nu = 1.4$\,GHz, which is the frequency band commonly used in radio observations, such as with the VLA \citep[see e.g.][]{XU03}. We also set $p = 2.2$, corresponding to electrons that are freshly accelerated by shocks \citep[see e.g.][]{ACH01}. Integrating the radio emissivity over the interval d$s$ along the line-of-sight gives the intensity 
\begin{equation}
\mathcal{I}_{\text R} 
\propto
\int {\rm{d}}s ~\mathcal{C} B_{\perp}^{(p+1)/2}\,,
\label{eq:IR}
\end{equation}
in erg\,s$^{-1}$\,cm$^{-2}$\,Hz$^{-1}$.

\begin{figure*}
 \begin{centering}
 \includegraphics[width=\textwidth,keepaspectratio=true]{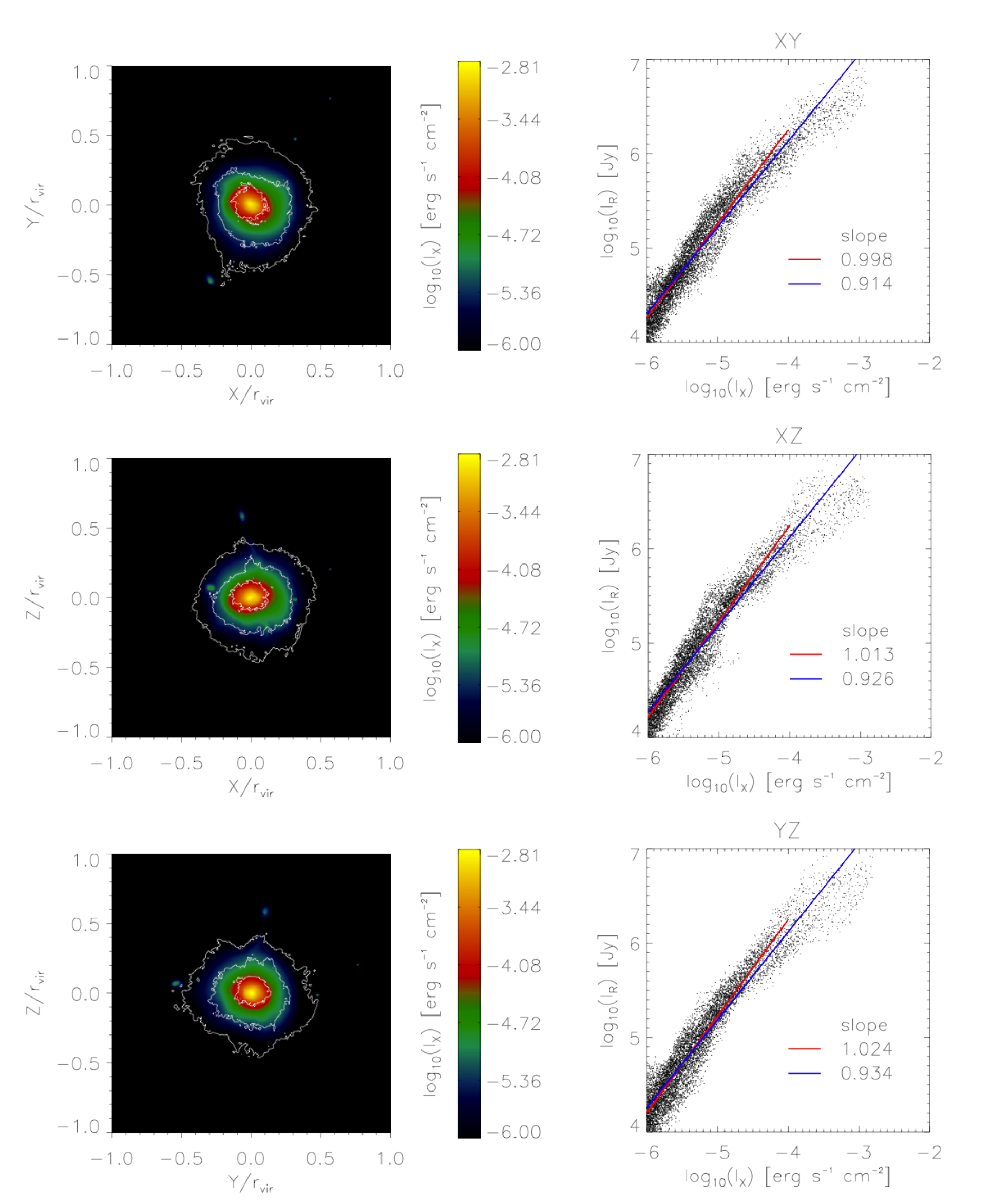}
 \caption{Left column: Contours of radio intensities $I_{\text R}$ at $10^{4}$, $10^{5}$ and $10^{6}$ $\rm{Jy}$, superimposed on the X-ray intensities $log_{10} I_{\text X}$ of the higher resolution cluster generated using the cleaned implementation, viewed in the XY (top), XZ (middle) and YZ (bottom) planes. Note that it is present-day configuration. The X-ray and radio intensities have similar morphologies. Right column: A point-by-point comparison of the X-ray and radio intensities in the XY, XZ and YZ planes, with a best-fit performed at $I_{\text X} \le 10^{-4}$ erg s$^{-1}$ cm$^{-2}$ (red) and across all intensities (blue). Clear to see that the correlation follows closely a power-law with a slope around unity.}
 \label{fig:GCobs}
 \end{centering}
\end{figure*}

The resulting X-ray and radio intensities are shown in the left panels of Fig. \ref{fig:GCobs}. We generate synthetic X-ray images of the cluster overlaid with radio contours in the $xy$, $xz$ and $yz$ planes. \replaced{The colourmap shows the total X-ray intensity, reaching a peak value of $I_{\text X} \approx 10^{-3}$\,erg\,s$^{-1}$\,cm$^{-2}$. We see some small disturbances in the X-ray morphology, which is consistent with the cluster experiencing a minor merger due to the infall of a small substructure}{The colours represent the total X-ray intensity on a logarithmic scale to improve visibility of the small $I_\text{X}$ values. We consider a minimum cut-off at $I_{\text X} = 10^{-6}$\,erg\,s$^{-1}$\,cm$^{-2}$ below which the X-ray emission would be too faint to be detected. From the X-ray images, we see irregular and asymmetric substructures, hinting that the cluster is likely to be disturbed and may have experienced a recent merger}. Superimposed over the X-ray images are synthetic radio contours in white, where the levels define intensities decreasing outwards at $10^{6}$, $10^{5}$ and $10^{4}\,\text{Jy}$, respectively. These values of intensities are independent of distance from the source and correspond to artificial fluxes of several $\mu$Jy per pixel at $1.4$\,GHz, which are consistent with various observations \citep[e.g.][]{GOV01A, MUR09, VAC10, MUR10, FAR13} and numerical simulations \citep[e.g.][]{VAC10, KOT11, GEN12, GOV13}. \footnote{Following the method by \citet{GEN12}, we estimate the expected flux arriving on Earth by multiplying the calculated total intensities with a factor $f = \pi r_{\text{beam}}^{2}/(4\pi D^{2})$, assuming $D\approx99\,\mathrm{Mpc}$ for the distance to the cluster \citep[e.g. Coma, see][]{REI02}, and treating $r_{\text{beam}}$ as the beam radius corresponding to an angular resolution of $\sim 5''$, if observed with, e.g. the VLA B-configuration at $1.4$\,GHz \citep[see e.g.][]{CON03}.} \replaced{We find that the X-ray and radio peaks overlap each other and lie on the potential minimum of the cluster}{As expected, the X-ray and radio peaks lie at the centre of the cluster with intensities declining with radius}.

Qualitatively, the X-ray images and radio contours show similar morphological features, indicating a spatial coincidence between the \replaced{electron density and the magnetic field}{thermal and relativistic components} in the cluster, \added{which is not surprising due to the magnetic field being effectively frozen in the gas}. To quantify this, we perform a local point-by-point comparison between the X-ray and radio intensities projected along the lines of sight: $x$, $y$ and $z$; adopting a similar approach to \citet{DOL01}. This method is unaffected by morphological differences and possible misalignments between the radio and X-ray maxima \citep{GOV01A}. \added{The simulated cluster has a bolometric X-ray luminosity of $L_{\mathrm{X},500}=3.11\times10^{44}$\,erg\,s$^{-1}$, a value that is marginally higher than observed \citep{Maughan2012} and likely impacted by the overcooling adiabatic simulations suffer from.} \replaced{W}{In so doing, w}e find a positive correlation between the radio and X-ray intensities in the cluster\footnote{We adhere to intensity in our calculations because it is independent of distance from the source. The conversion between intensity and flux/surface brightness does not affect the value of the best-fit slope.}, which appears to satisfy a power-law relation
\begin{equation}
\mathcal{I}_{\rm R} = A\,I_{\rm X}^{b} \:,
\end{equation}
where $A$ is the normalization constant and $b$ is the slope of the $\mathcal{I}_{\text R}-I_{\text X}$ relation. We therefore use the method of least squares to fit the linear relation
\begin{equation}
\log_{10}(\mathcal{I}_{\mathrm{R}}) = \log_{10}(A)+ b\,\log_{10}(I_{\mathrm{X}})\:,
\end{equation}
and the results are shown in the right panels of Fig. \ref{fig:GCobs}. Clearly, there is some scatter around the lines of best-fit, indicating that \replaced{the correspondence between magnetic field and electron density is not one-to-one}{the cluster is dynamically disturbed}. We further investigate the \replaced{impact of the adiabatic nature of the simulation}{effects} on the scaling relation by performing a best-fit across all data (blue line) and compar\replaced{ing}{e} it to that where the central region of the cluster is excised (red line). The latter omits the anomalous, over-dense X-ray bright region of the cluster, which is a consequence of the simulation \deleted{being adiabatic and}lacking \deleted{of} additional physics, such as radiative cooling and AGN feedback. By setting the upper limit at $I_{\text X} = 10^{-4}$ erg s$^{-1}$ cm$^{-2}$, we calculate the best-fit slopes, obtaining $0.998$, $1.013$ and $1.024$ along the $z$, $y$ and $x$ directions, respectively. These values are slightly \replaced{steeper}{larger} than those derived from an all-data fit: $0.914$, $0.926$ and $0.934$; \replaced{demonstrating}{implying} that the centre-excised $\mathcal{I}_{\text R}-I_{\text X}$ relation is \replaced{closer to unity}{tighter}, and that the central region of the cluster \replaced{is impacted by the missing physics}{does not follow a universal behaviour}. \added{A power-law close to unity is similar to observed values, even though the cluster is significantly smaller than those with observed haloes, which suggests that it arises naturally from the magnetic field being frozen into the gas} \citep{FER01, GOV01A}. \footnote{It is worth noting that, while $\int_{-1}^{1} dx~f(x) \propto \int_{-1}^{1} dx~f(\sqrt{1-x^2})$, $f(x)$ is not necessarily proportional to $f(\sqrt{1-x^2})$.}

It is worth mentioning a few caveats. Whilst the magnetic fields are relatively weak and frozen into the hot plasma, their interactions with the relativistic plasma is not known from our simulations. We therefore have to make simplifying assumptions whereby we scale the relativistic energy density with the thermal energy density, as well as treat the energy spectrum of the relativistic electrons as a power law. As a result, the spatial coincidence between the radio halo emission and the hot gas regions is somewhat expected. \added{The result of the \citet{Planck2013} suggest that correlation between the radio emission and, in their case, the SZ signal could be used to distinguish between re-acceleration mechanisms or the distribution of the magnetic field, which may hint at its origin. Although we find a linear relation between the X-ray and radio emission, we are unable to comment of these results because our model for non-thermal electrons is too crude, and a study of the impact of different acceleration mechanisms is beyond the scope of this paper.}

\section{Conclusions}
\label{sec:concs}
In this paper we have implemented the recently proposed artificial resistivity switch of \citet{TRI13} and a hyperbolic cleaning scheme \citep{TRI12} in to the SPMHD code \textsc{gcmhd+}. Using standard idealised MHD test cases we have examined the impact of these additions. We found that the new resistivity switch tracks discontinuities in the magnetic field significantly better then the previous resistivity switch of \citet{PRI05}. It applies more resistivity where it required and significantly less away from discontinuities. However, the increase in applied resistivity leads to some small scale features being smoothed out completely. The inclusion of the hyperbolic cleaning scheme leads to a significant reduction in the divergence error throughout the test simulations with negligible impact on the result produced by the code.

We then used six different configurations of \textsc{gcmhd+} to perform a zoom simulation of the formation of a galaxy cluster with a primordial seed magnetic field embedded in the gas particles. We found that the choice of numerical MHD scheme has a significant impact on the amplitude and topology of the final cluster magnetic field. The configurations that did not include the hyperbolic cleaning scheme produced a magnetic field with a significantly greater amplitude, but the stronger field was associated with larger divergence errors. The inclusion of the hyperbolic cleaning scheme produced a significant reduction in the divergence error throughout the cluster volume and maintained the divergence of the magnetic field to a few percent of the total field amplitude. \added{The inclusion of resistivity resulted in a reduction of the magnetic field amplitude, increment with previous numerical work} \citep[e.g.][]{DOL09,BON11}. We found the that the TP13 resistivity switch was too strong for cosmological simulations and lead to significant suppression of the cluster magnetic field amplitude.

Following the evolution of the seed field with redshift we found that failing to suppress the growth of the divergence of the magnetic field lead to a steeper rate of increase of average magnetic energy density within the cluster volume. This is due to both the true magnetic field and the divergence of the magnetic field being amplified by the turbulent motions of the ICM. Therefore, to ensure that the evolution of the seed magnetic field is more reliably captured, the growth of the divergence of the magnetic field should be suppressed, for example using a hyperbolic cleaning scheme. 

The cleaned and PM05 cleaned configurations, which were found to suppress the growth of the numerical divergence and did not applied excessive resistivity, were then used to simulate a higher resolution version of the simulations. The increase in resolution lead to the PM05 resistivity switch applying significantly more resistivity throughout the cluster volume and this suppressed the amplitude of the cluster magnetic field. The cleaned implementation produced a galaxy cluster with a $\mu\text{G}$ amplitude magnetic field that steeply declined in amplitude with radius, which is in good agreement with the observations. We then produced synthetic emission predictions for the cluster simulated with the cleaned implementation. We calculate the radio and X-ray intensities of the cluster and find that they possess similar morphological features, which are in agreement with observations of several massive, merging clusters. Finally, we perform a local point-by-point comparison of the radio and X-ray images of the simulated cluster. A power law relation is found, with a best-fit slope of $b\sim1$, which is consistent with the values obtained for observed galaxy clusters.

Our results indicate that the hyperbolic cleaning scheme should be applied in current state of the art SPMHD simulations. However, the artificial resistivity scheme should be carefully chosen based on the astrophysical phenomena of interest. Considering the success of \citet{TRI13} in the strong magnetic field regime, we expect that artificial resistivity schemes will need to be calibrated for cosmological MHD simulations.

\section*{Acknowledgements}

The authors acknowledge the support of the UK's Science \& Technology Facilities Council (STFC Grant ST/H00260X/1). A.\,Y.\,L. On gratefully acknowledges support from the Government of Brunei under the Ministry of Education Scholarship. The calculations for this paper were performed on Cray XT4 at Centre for Computational Astrophysics, CfCA, of National Astronomical Observatory of Japan and the DiRAC Facility jointly funded by STFC and the Large Facilities Capital Fund of BIS. The authors acknowledge support of the STFC funded Miracle Consortium (part of the DiRAC facility) in providing access to the UCL Legion High Performance Computing Facility. The authors additionally acknowledge the support of UCL's Research Computing team with the use of the Legion facility. We also thank James M. Stone for making \textsc{athena}, which we used to obtain reference solutions to many of the tests, publicly available.

\bibliographystyle{mnras}
\bibliography{main}

\appendix
\section{Shock tube 1B}
\label{sec:AppST}
\begin{figure}
 \begin{centering}
 \includegraphics[width=\columnwidth,keepaspectratio=true]{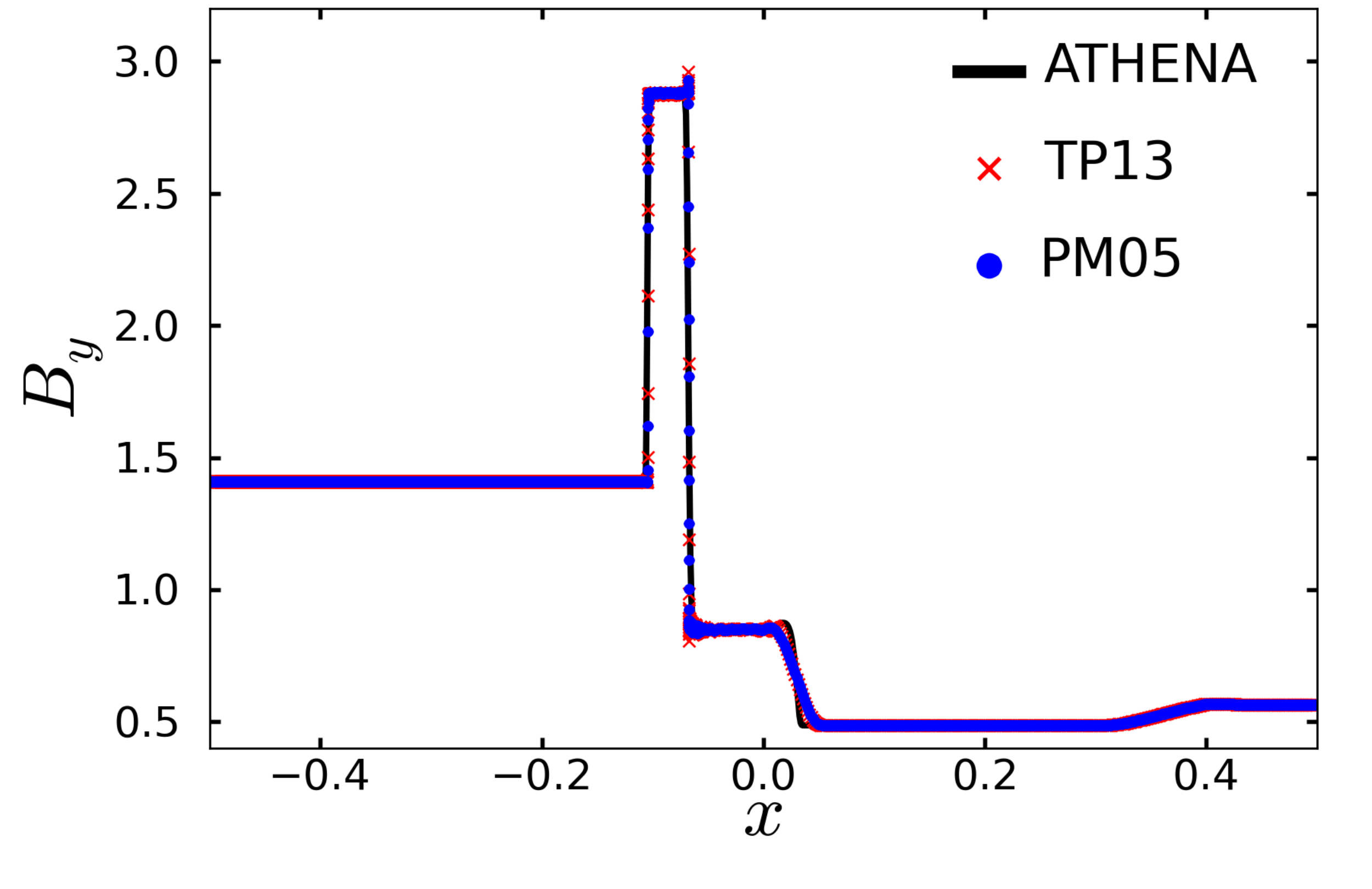}
 \caption{The $y$ component of the magnetic field for shock tube test 1B at $t=0.03$. We plot the reference \textsc{athena} (black line), the PM05 implementation (blue circles) and TP13 implementation (red crosses) results.}
 \label{fig:1B}
 \end{centering}
\end{figure}

\added{To test the impact of different configurations we performed a suite of idealized test simulations. The one dimensional shock tube tests demonstrate the ability of the code to capture shock and rarefaction waves in both the strong and weak regimes. Due to their $1\mathrm{D}$ nature the divergence of the magnetic field is automatically conserved. However, they provide a good test of the ability of the artificial resistivity switch to capture discontinuities in the magnetic field. In Fig. \ref{fig:1B} we show the magnetic field in the $y$ direction, $B_{y}$, as a function of position for the shock tube 1B test at $t=0.03$. We plot the results produced by the 'PM05 cleaned' and 'TP13 cleaned' implementations against the result produced by \textsc{athena}. The two implementations are in reasonable agreement with the \textsc{athena} result, with a small over-shoot and some noise at $x=-0.05$. The PM05 cleaned implementation produces slightly less noise at $x=-0.05$ and a smaller over-shoot. Both implementations smooth the feature at $x=0.025$ more than the ATHEAN result.}

\bsp	
\label{lastpage}
\end{document}